\documentclass[aps,prl,twocolumn,superscriptaddress,longbibliography]{revtex4-2}

\usepackage{mathrsfs}
\usepackage{tabularx}
\usepackage{slashed}
\usepackage{amsmath}
\usepackage{amsfonts}
\usepackage{graphicx,color}
\usepackage{times}
\usepackage[caption=false]{subfig}
\usepackage[colorlinks,linkcolor=red,citecolor=blue]{hyperref}
\newcommand{\comment}[1]{}
\AtBeginDocument{%
    \newwrite\bibnotes
    \def\bibnotesext{Notes.bib}
    \immediate\openout\bibnotes=\jobname\bibnotesext
    \immediate\write\bibnotes{@CONTROL{REVTEX41Control}}
    \immediate\write\bibnotes{@CONTROL{%
    apsrev42Control,author="08",editor="1",pages="1",title="0",year="1"}}
     \if@filesw
     \immediate\write\@auxout{\string\citation{apsrev42Control}}%
    \fi
}%

\begin{document}

\title{Non-Abelian Fractional Chern Insulators and Competing States in Flat Moir\'e Bands}
\author{Hui Liu}\thanks{hui.liu@fysik.su.se}
\affiliation{Department of Physics, Stockholm University, AlbaNova University Center, 106 91 Stockholm, Sweden}

\author{Zhao Liu}\thanks{zhaol@zju.edu.cn}
\affiliation{Zhejiang Institute of Modern Physics, Zhejiang University, Hangzhou 310058, China}
\affiliation{Zhejiang Key Laboratory of Micro-Nano Quantum Chips and Quantum Control, School of Physics, Zhejiang University, Hangzhou 310027, China}

\author{Emil J. Bergholtz}\thanks{emil.bergholtz@fysik.su.se}
\affiliation{Department of Physics, Stockholm University, AlbaNova University Center, 106 91 Stockholm, Sweden}

\date{\today}

\begin{abstract}
Breakthrough experiments have recently realized fractional Chern insulators (FCIs) in moiré materials. However, all states observed are Abelian, the possible existence of more exotic non-Abelian FCIs remains controversial both experimentally and theoretically. Here, we investigate the competition between charge density wave (CDW) order, gapless composite fermion liquid (CFL), and non-Abelian Moore-Read states at half-filling of a moiré band. Although groundstate (quasi-)degeneracies and spectral flow are not sufficient for distinguishing between charge order and Moore-Read states, we find evidence using entanglement spectroscopy that both these states of matter can be realized with Coulomb interactions. By further analyzing the graviton excitations of Moore-Read states, we unveil that the ground states exhibit a mixed behavior of Pfaffian and anti-Pfaffian, despite the weak breaking of particle-hole symmetry. In a double twisted bilayer graphene model, transitions between these phases can be driven by the coupling strength between the layers: at weak coupling there is a CFL phase and at strong coupling a CDW order emerges. Remarkably, however, there is compelling evidence for a non-Abelian Moore-Read FCI phase at intermediate coupling.      

\end{abstract}

\maketitle
\emph{Introduction.} --- 
Moiré materials \cite{andrei2020graphene, Andrei_review, caoUnconventionalSuperconductivityMagicangle2018, Cao2018, Bistritzer2011} have provided a long sought platform for realizing fractional Chern insulators (FCIs) \cite{LIU2024515}:
recently theoretically predicted and subsequently experimentally observed both in graphene based heterostructures~\cite{PhysRevLett.124.106803,repellinChernBandsTwisted2020,PhysRevResearch.2.023237,zhaoTDBG,xie2021fractional,lu2023fractional,spanton2018observation} and twisted transition metal dichalcogenide bilayers~\cite{PhysRevResearch.3.L032070,PhysRevB.107.L201109,FCI_MoTe2_1,FCI_MoTe2_2,FCI_MoTe2_3, PhysRevX.13.031037}.
 While remarkable in not requiring any external magnetic field \cite{FCI_MoTe2_1,FCI_MoTe2_2,FCI_MoTe2_3, PhysRevX.13.031037,lu2023fractional} and opening up the prospects of high temperature anyon physics, all states observed thus far Abelian. Non-Abelian states, which would entail an even richer phenomenology and prospects for applications \cite{Nayak2008}, remain elusive in experiments and their theoretical identification is subject to intense discussion.

Even in FCI toy models, where Abelian FCIs are ubiquitous
\cite{bergholtz2013topological,PARAMESWARAN2013816,kolread,tang_high-temperature_2011,PhysRevLett.106.236803,PhysRevLett.106.236804,sheng2011fractional,PhysRevX.1.021014,mollercooper,PhysRevLett.105.215303}, non-Abelian states are notoriously hard to stabilize. In fact, most studies of non-Abelian FCIs feature rather artificial multi-body interactions \cite{Wang2012,Zoology2012,Repellin2013,Wu2013,Bergholtz2015,Behrmann2016}, strategically chosen to mimic their multi-body pseudopotential parent Hamiltonians that may be derived from the operator product expansion of the pertinent conformal field theories \cite{MOORE1991362,greiter1992,read_rezayi}. With more realistic two-body interactions it is much harder to stabilize non-Abelian FCIs and only a few models are known to harbour such \cite{NAlong2013,Wang2015}. In the context of Landau levels there is evidence of the paradigmatic non-Abelian fractional quantum Hall states, namely the Moore-Read (MR) states \cite{MOORE1991362}, in a narrow range of parameter space that fortunately includes the second Landau level \cite{Morf1998,Rezayi2000}.

\begin{figure}[t]
\centering
\includegraphics[width=0.9\linewidth]{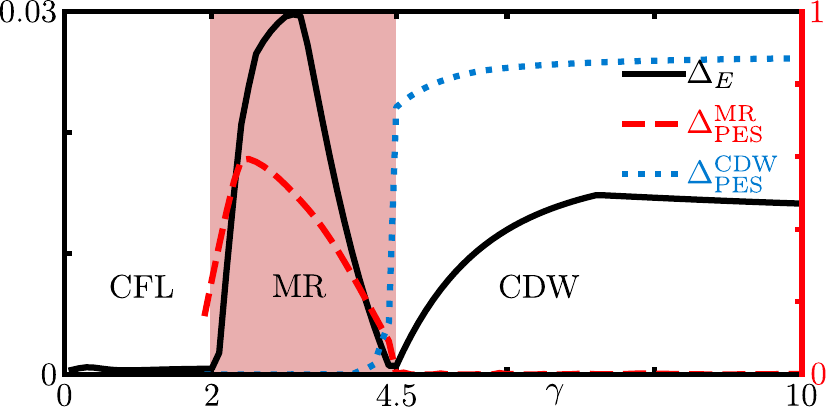}
\caption{{\bf Competition and phase transition between Moore-Read states and CDWs}. Particle-cut entanglement spectrum gap $\Delta_{\text{PES}}$ for quasihole counting of the CDW (blue dotted line) and MR states (red dashdot line), along with the energy difference between the MR ground states and the first excitation state (black solid line) $\Delta_E$, as a function of the coupling strength $\gamma$ (eV), respectively. Here we use a system size with $26$site-clusters at half-filling, and $N_A=4$ for subsystem $A$ in particle-cut entanglement spectrum calculations, and a bare Coulomb interaction.}
\label{fig:transition}
\end{figure}

While clearly extraordinarily challenging to realize in a lattice setting and without a strong magnetic field, very recent experiments claiming to show signs of a fractional topological insulator in twisted semiconductor tMoTe$_2$ \cite{non_Abelian_experiment} have reignited the interest in non-Abelian moiré based FCIs \cite{mr_Aidan,2024arXiv240300856F, mr_Wang, mr_Ahn, mr_Xu,mr_zhang,mr_Chen,yahui_zhang_non_abelian, Inti_sodemann_fractional_quantum_spin_Hall}. The evidence for non-Abelian FCIs in these systems is intriguing: quantum numbers and quasi-degenerecies match those of the MR states \cite{mr_Aidan,mr_Xu} and bandstructures show similarities with the second Landau level \cite{2024arXiv240300856F}.
However, charge density waves (CDWs) \cite{Wilhelm2021,Wilhelm_2023} and composite Fermi liquids (CFLs) \cite{PhysRevLett.131.136501, PhysRevLett.131.136502} are fierce competitors to the non-Abelian FCIs. In fact, for some sample geometries all these may have the same quasi-degeneracies and (momentum) quantum numbers. In particular, the CDWs may have the same ground-state quasi-degeneracies and occur in the same momentum sectors for any sample size, making them very hard to distinguish from the sought after MR FCIs. This is in particular pressing given the relatively small system sizes available to numerical simulation for which charge order may be less developed due to finite size effects and/or in the case when the microscopic nature of the competing local order is not a priori obvious. In this case a more careful investigation of entanglement spectrum is a powerful method for distinguishing the competing phases \cite{PhysRevX.1.021014}.

\begin{figure*}
\centering
\includegraphics[width=\linewidth]{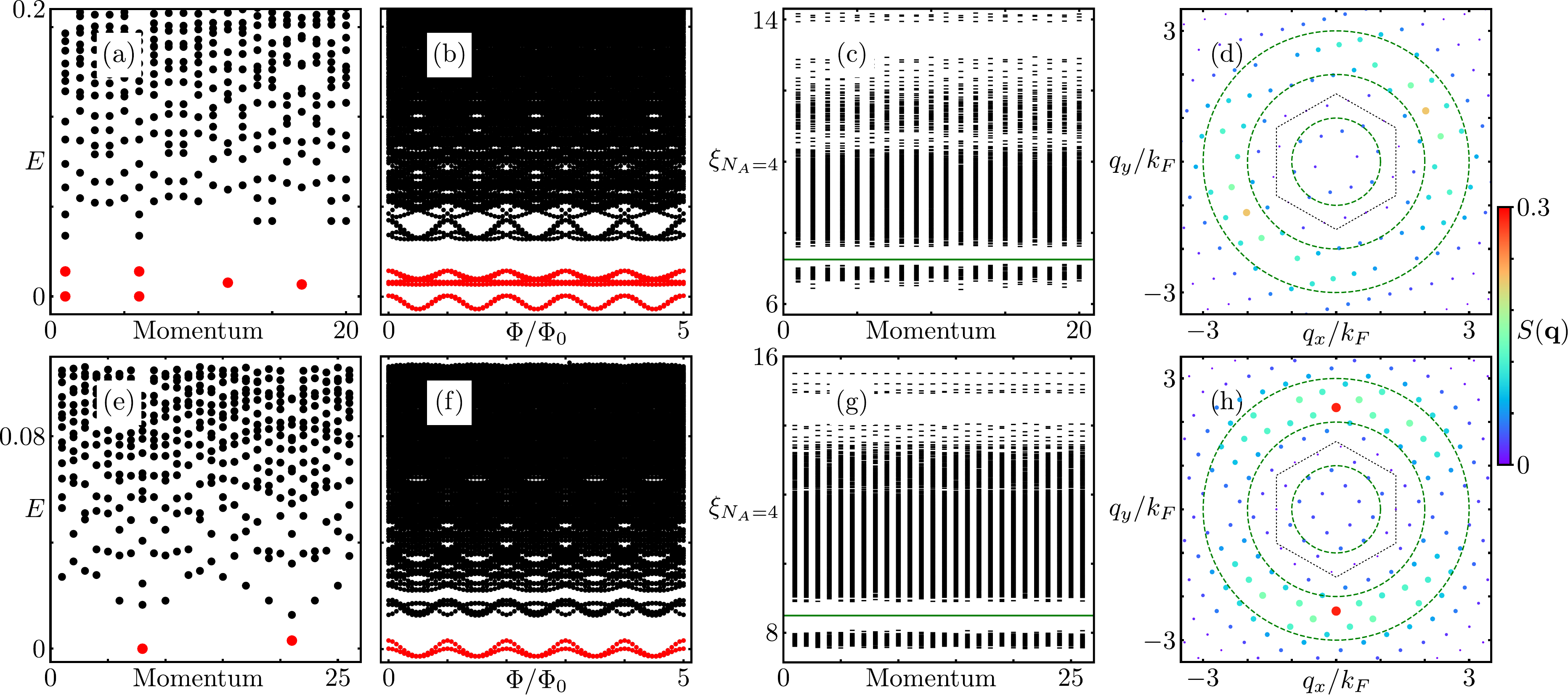}
\caption{{\bf Evidence for charge density waves}. Panels (a-d) are the low-lying energy spectrum, spectral flow, particle-cut entanglement spectrum with $N_A=4$, and the structure factor for the $20$site-cluster system at half filling (even number of electrons). 
Panels (e-h) present the corresponding results for the $26$site-cluster system at half filling (odd number of electrons). 
In the energy spectrum, the red dots indicate the ground states satisfying the center-of-mass momentum of Moore-Read states. 
In the entanglement spectrum, the number of states below the first entanglement gap (the green solid line) is $420$ for $20$site-cluster and $1430$ for $26$site-cluster, which adheres to the counting rule for CDWs. In the structure factor, the solid line indicates the moiré BZ and $k_F$ represent the Fermi wave vector.
For all plots, we employ $\gamma=5.5 \ {\rm eV}$ and a bare Coulomb interaction.
\label{fig:CDW}
}
\end{figure*}

\begin{figure*}
\centering
\includegraphics[width=\linewidth]{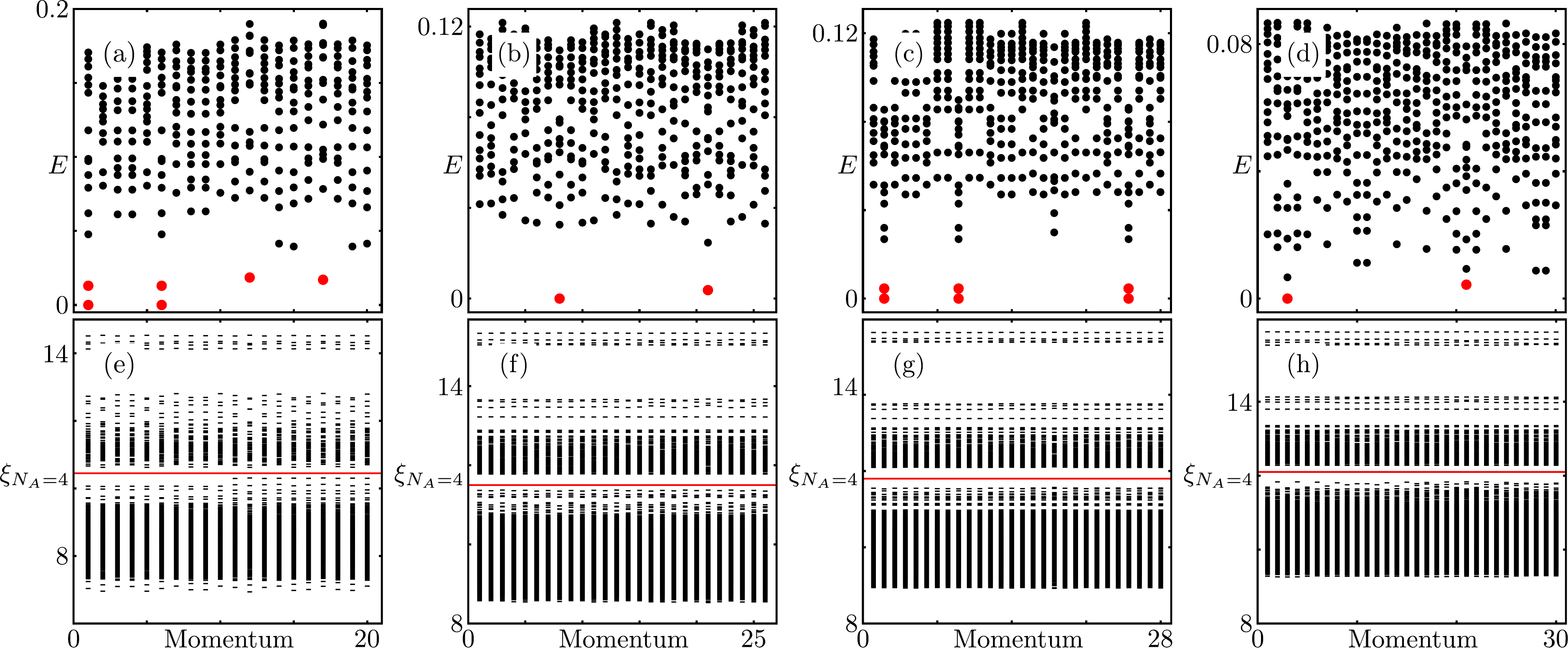}
\caption{{\bf Evidence for Moore-Read states}. Panels (a-d) are the low-lying energy spectrum for system with $20$, $26$, $28$, and $30$site clusters, respectively. Here, the red dots represent the Moore-Read states. Panels (e-h) are their corresponding particle-cut entanglement spectrum for $N_A=4$. The number of states below the first entanglement gap (the red solid line) is $3965$, $13338$, $18571$, and $25185$, matching the quasihole counting for Moore-Read states. For all plots, we use $\gamma=3.5 \ {\rm eV}$.
\label{fig:MR}
}
\end{figure*}

Here we use entanglement spectroscopy to demonstrate that non-Abelian MR FCIs can indeed be realized in moiré systems. Our investigation also reveals that there is a fierce competition between all three aforementioned phases of matter and novel phase transitions may occur without changes in ground states degeneracies. This is shown in Fig. \ref{fig:transition}: in a simple model of a double twisted bilayer graphene \cite{2024arXiv240300856F}, changing the interlayer coupling tunes the system from MR FCIs to low-entanglement states that exhibit the entanglement properties expected for CDWs. Both these phases exhibit the salient energetic characteristics of the MR states, yet are separated by an energy gap closing and with two tell-tale entanglement gaps opening and closing, respectively. Below we detail the setup followed by a description of evidence for the CDW (Fig. \ref{fig:CDW}) and MR (Fig.~\ref{fig:MR}) phases (as well as the transition between them in the end matter). We also briefly comment on CFLs and put our results into a broader context.  


\emph{Setup.} --- 
\label{sec:setup}
We consider a valley and spin polarized Hamiltonian describing double sheets of twisted bilayer graphene (TBG) of the form~\cite{2024arXiv240300856F},
\begin{eqnarray}
    H(\bf r)=\begin{pmatrix}
        u_1&D_1^\dagger(\bf r)&0&0\\
        D_1(\bf r)&u_2&\gamma I&0\\
        0&\gamma I&u_3&D_2^\dagger(\bf r)\\
        0&0&D_2(\bf r)&u_4
    \end{pmatrix}.
\end{eqnarray}
Here, the two $2\times 2$ diagonal blocks represent the two single-sheet TBGs with
\begin{eqnarray}
    D_{i}(\bf r)=\begin{pmatrix}
    -2ik_\theta^{-1}\Bar{\partial}&\alpha U(\bf r)\\
    \alpha U(-\bf r)&-2ik_\theta^{-1}\Bar{\partial}
\end{pmatrix}.
\end{eqnarray}
$\gamma$ controls the interlayer coupling strength between the twisted bilayers,
$u_{j}$ is the gate potential, $k_\theta=4\pi/3a_M$ with $a_M$ the moir\'e length, $\Bar{\partial}=\frac{1}{2}(\partial_x+i\partial_y)$, and $U(\mathbf{r})=e^{-i\mathbf{q}_1\cdot\bf r}+e^{i\phi}e^{-i\mathbf{q}_2\cdot\bf r}+e^{-i\phi}e^{-i\mathbf{q}_3\cdot\bf r}$. 
Setting $\gamma=0$ leads to two decoupled copies of chiral twisted bilayer graphene (cTBG)~\cite{PhysRevLett.122.106405, PhysRevResearch.2.023237}.
For the results present in the main text, we use the first magic angle for the cTBG ($\alpha\approx 0.586$) and $u_1=50 \ {\rm meV}$, $u_2=u_3=u_4=0$ to separate the middle flat bands (see the band dispersion in the supplementary material (SM)~\cite{SupMat}).


It has been recently suggested that coupling multiple TBG layers might enable the flat band above the charge-neutrality point to mimic higher Landau level (LL) physics~\cite{2024arXiv240300856F}, which then offers a potential for realizing non-Abelian topological states at fractional band filling \cite{Morf1998,Rezayi2000}.
For this purpose, we now consider interactions in this band. 
To begin, we first assess the tendency toward valley polarization by projecting the interaction into the flat band above the charge-neutrality point, including both $K$ and $K'$ valleys. In the parameter regime of interest, we indeed find valley polarization across different system sizes at half band filling, where the low-lying energy states fully occupy at either $K$ or $K'$ valley  (see the SM~\cite{SupMat}), which justifies focusing only on the flat band above the charge-neutrality point of the $K$ valley.
We also neglect the contribution of the band dispersion, which is well motivated in the strong interaction limit. While in realistic systems effects of band mixing may be substantial \cite{xu2024maximally,yu2024fractional,abouelkomsan2024band}, we are here mainly interested in the fundamental question of realizing novel states within a given flat band.
Throughout, we use the bare Coulomb interaction. 

We focus on the Moore-Read states, a well-known non-Abelian topological phase that emerges at $1/2$ band filling, first proposed by Moore and Read~\cite{MOORE1991362}. Their momenta can be identified by considering the thin-torus (aka root) configurations. For MR states, these configurations have two distinct types in Fock space, $11001100\cdots$ and $10101010\cdots$, respectively \cite{Bergholtz2006,Seidel2006, Ardonne_2008,PhysRevB.73.245334}, which can be derived by considering the thin-torus limit of a three-body parent repulsion leading to the selection rule that there are no more than two particles on any four adjacent sites in the one-dimensional representation. Combining their center-of-mass translations, two different ground-state degeneracies appear: a sixfold degeneracy for an even number of electrons and a twofold degeneracy for an odd number of electrons (since only the $10101010\cdots$ states are compatible with the periodic boundary conditions). 
We emphasize that the competitors of the MR states may occur in the same momentum sectors.


\emph{Charge density wave evidence.} --- 
\label{sec:CDW} 
We first consider strong interlayer coupling in the double TBGs model. In this regime, we find two distinct types of ground-state degeneracies, as exemplified in Figs.~\ref{fig:CDW}(a) and \ref{fig:CDW}(e) for $\gamma=5.5 \ {\rm eV}$. In our finite-size samples, both the fold of degeneracy and the ground-state momenta match the predictions by the aforementioned exclusion rule of the MR states. By applying an external magnetic flux, we further confirm the presence of an energy gap between the ground states and the excited states, below which the sixfold (even number of electrons) and twofold (odd number of electrons) quasi-degenerate ground states flow into each other [see Figs.~\ref{fig:CDW}(b) and \ref{fig:CDW}(d)]. The total quantum metric $\mathcal{G}\equiv \frac{1}{2\pi}\int_{\rm MBZ}{\rm tr}\ g({\bf k})d^2{\bf k}$ of the flat band, which is commonly used as an indicator of the closeness to a Landau level, indeed approaches $3$ -- the value of the first Landau level (1LL), with increasing interlayer coupling [$g({\bf k})$ is the quantum metric]. These findings, aligned with recent numerical observation in tMoTe$_2$~\cite{mr_Aidan, mr_Wang, mr_Ahn, mr_Xu, mr_zhang}, seemingly suggest the existence of non-Abelian MR states with strong interlayer coupling. 
 
Surprisingly, upon a particle-cut entanglement spectrum (PES) analysis~\cite{li_entanglement_2008, sterdyniak_extracting_2011, PhysRevX.1.021014} (see the definition in the SM~\cite{SupMat}), we find that the ground states instead exhibit characteristics of the CDW. 
In Figs.~\ref{fig:CDW}(c) and \ref{fig:CDW}(g), a narrow band is discernible at the bottom of the PES, separated from the higher entanglement levels by a visible gap. 
Remarkably, in both cases, the number of states below the gap is very different from the quasihole counting for non-Abelian MR states: there are many fewer eigenvalues below the gap. Instead, the observed counting matches a CDW order, where the number of states below the gap is just the way of putting $N_A$ particles into $N_e$ specific positions~\cite{2012arXiv1204.5682B}. Moreover, a visible gap at the counting of MR states is lacking, which provides corroborating evidence against such states.
Therefore, even though system-size-dependent ground-state degeneracies and a gapped spectral flow are fulfilled, the presence of MR states is still not guaranteed due to the competing order. 

Motivated by the PES findings, we further employ calculations of the ground-state structure factor $S({\bf q})$, which is a typical method for validating the charge order [see Figs.~\ref{fig:CDW}(d) and \ref{fig:CDW}(h)]. Indeed, we find pronounced peaks in $S({\bf q})$ for both system sizes, in particular the one with odd number of electrons (26site-cluster). This supports the existence of charge order at strong interlayer coupling. Similar peaks in $S({\bf q})$ together with the corresponding PES gap for CDW ground states persist for larger systems, as shown in the SM~\cite{SupMat}. 




\emph{Moore-Read states.} ---
\label{sec:MR} 
Moving to intermediate coupling strength, we find compelling evidence for the true non-Abelian MR states. As shown in Figs.~\ref{fig:MR}(a)-(d) for $\gamma=3.5 \ {\rm eV}$, the low-lying energy spectrum for various system sizes exhibits correct ground-state degeneracies and center-of-mass momenta, as predicted by the exclusion rule for the MR states. Importantly, these ground states remain well isolated from the excited states by a visible gap, which persists as well upon inserting flux through the handles of the torus [see the SM~\cite{SupMat}].

Next, to eliminate the possibility of charge orders, we again examine the PES of the sixfold ground states (for even number of electrons) and the twofold ground states (for odd number of electrons). In Figs.~\ref{fig:MR}(e)-\ref{fig:MR}(h), a clear entanglement gap consistently appears between the bottom wide band and the middle narrower band. 
Remarkably, the number of states below the gap precisely matches the quasihole counting of the MR states. The persistence of the PES gap across various system sizes and different $N_A$ serves as a hallmark of the robustness of the MR states in moiré minibands. In fact, such strong evidence for the MR states does not even exist in the 1LL. 


Now we conclude that the MR states only appear at intermediate coupling strength instead of at (infinitely) large $\gamma$, which was often thought as the optimal point to capture the 1LL physics as $\mathcal{G}=3$. We attribute this discrepancy to the role of quantum metric fluctuation in determining the underlying physics. The 1LL has both $\mathcal{G}=3$ and uniform quantum metric. By contrast, a Chern band cannot possess both. A compromise between approaching $\mathcal{G}=3$ and minimizing quantum metric fluctuations must be made in a Chern band to realize the 1LL physics. In line with this, we indeed find both the minimal metric fluctuation and $\mathcal{G}$ not too far from $3$ at intermediate coupling, where the MR states are favored [see the SM~\cite{SupMat}]. 
We note that although the minimum of quantum metric fluctuation lies within the MR state regime, it is difficult to establish the correspondence between the phase transition point---identified via the energy spectrum or PES---and a strict threshold in quantum metric and its fluctuations, as these quantities are purely single-particle indicators and cannot fully determine the many-body physics, which also depends on interaction details.


After confirming the existence of MR states in the moiré miniband, a natural question arises: are these non-Abelian MR states Pfaffian or anti-Pfaffian? These two possibilities are particle-hole conjugate of each other and share the same quasihole excitation statistics, making it necessary to discuss their distinctions in the context of moiré systems. In LLs, the presence of particle-hole symmetry gives rise to cat states of Pfaffian and anti-Pfaffian. Then, a total of $12$ ground states are expected in the thermodynamic limit. In the case of the double TBG model, the absence of magnetic translation symmetry leads to particle-hole symmetry breaking. Under the particle-hole transformation $c^\dagger_{\bf{k}}\rightarrow d_{\bf{k}}$, a single-hole kinetic energy term $H_{\text{single-hole}}=\sum_\mathbf{k}E_{\text{h}}(\mathbf{k})n(\mathbf{k})$ with $E_{\text{h}}(\mathbf{k})= \sum_{\mathbf{k}'\in \text{MBZ}} \big( V_{\mathbf{k}'\mathbf{k}\mathbf{k}'\mathbf{k}} +  V_{\mathbf{k}\mathbf{k}'\mathbf{k}\mathbf{k}'} - V_{\mathbf{k}\mathbf{k}'\mathbf{k}'\mathbf{k}}-V_{\mathbf{k}'\mathbf{k}\mathbf{k}\mathbf{k}'}\big)$ appears. In general, the hole dispersion $E_{\text{h}}(\mathbf{k})$ varies with ${\bf k}$ for moiré flatbands (see the SM~\cite{SupMat}, which includes Ref.~\cite{Yoshioka1983}). Furthermore, the occupation number of ground states is not constant as well (which usually exhibits a center-of-mass dependence). Together, these factors suggest that the single-hole kinetic term leads to an energy difference between Pfaffian and anti-Pfaffian states; thus, one of them may be selected out. However, as non-constant $E_{\text{h}}(\mathbf{k})$ (depending on the fluctuation of quantum metric~\cite{PhysRevResearch.5.L012015,2024arXiv240908324J}) in this case is nearly uniformly distributed in MBZ (as shown in the SM~\cite{SupMat}), we expect that the effect of $H_{\text{single-hole}}$ becomes significant only in large systems with sufficintly many accessible ${\bf k}$ points~\cite{liu2024theorygeneralizedlandaulevels}. This makes the observation of state selection very challenging in small systems within the capability of numerical simulation. 

The mixed nature of Pfaffian and anti-Pfaffian states can be further verified by studying the chirality of their graviton excitations~\cite{form_factor_ll, graviton0,graviton00, graviton000}, which has been recently generalized to moiré systems~\cite{graviton1, graviton2, graviton3}. Using graviton operators respecting the crystalline symmetry of the moiré lattice, we found that for both even and odd number of electrons, the corresponding graviton spectra carry nearly equal weights of positive and negative chiralities, indicating a superposition of Pfaffian and anti-Pfaffian states (see more details in the SM~\cite{SupMat}). 
Such a phenomenon has also been observed in the first LL, where exact particle-hole symmetry enforces identical graviton spectra for both chiralities~\cite{moore_read_graviton}. In the present case, particle-hole symmetry is only weakly broken, consistent with the nearly uniform single-hole energy. To suppress one of the two chiralities, an alternative approach is to include three-body interactions (arising from band mixing effects), which explicitly break particle-hole symmetry. Then, one of Pfaffian or anti-Pfaffian states may be selected out, depending on the sign of such interactions. A systematic exploration of such a phenomenon in non-Abelian moiré FCIs warrants future verification.

\emph{Composite Fermi liquids.} ---
A third competing phase in these systems is comprised by gapless CFLs. 
It originally occurs in the lowest LL at $1/2$ or $1/4$ band filling~\cite{Jain_cfl}. 
Within our system, this phenomenon arises at weak coupling strength $\gamma\rightarrow 0$, where the full system decouples into two independent chiral TBGs, each resembling the characteristic behaviors of the original cTBG, which has been verified that its flat band captures the lowest LL physics~\cite{PhysRevResearch.2.023237}. 
This then offers a convincing explanation for the emergence of gapless CFLs. 
We confirm this by both the low-lying energy spectrum, the electron occupation of the CFL ground states, and the static structure factor, which aligns with the recent reports of CFLs in tMoTe$_2$~\cite{PhysRevLett.131.136501, PhysRevLett.131.136502}. In particular, the structure factor becomes singular around $|\mathbf{q}|=2k_F$, due to the backscattering process of composite fermions at the Fermi surface of $\mathbf{k}=k_F$ [see the SM~\cite{SupMat}, which includes Ref.~\cite{composite_fermion_science}].

\emph{Discussion.} ---
In this work we have strongly corroborated recent claims that non-Abelian MR states may be realized in moiré systems. At the same time our detailed entanglement spectroscopy together with the structure factor highlights the fact that ground-state degeneracies and momentum quantum numbers alone are insufficient for establishing the presence of this exotic phase of matter. In particular our analysis demonstrates that competing CDW order can be the underlying reason for states that at first glance appear to be MR states. 

We then uncover genuine MR states in the intermediate coupling strength regime, where these states isolated from excited states consistently emerge across various system sizes and are accompanied by the correct quasihole excitations. 
Moreover, the Pfaffian and anti-Pfaffian nature of these states is verified by their geometric excitations, in which the nearly identical spectral weights suggest a mixed Pfaffian and anti-Pfaffian character, due to the weakly breaking of particle-hole symmetry. 

By tuning the coupling strength between two sheets of TBGs, we further identify the possibility for a direct phase transition between the charge density wave and the Moore-Read states, where the entanglement spectrum exhibits a consistent behavior with the energy gap closing and reopening.
 
Our work serves as a prototypical example of the interplay between Moore-Read states and competing states in moir\'e systems, offering compelling avenues for future research. 
Firstly, while non-Abelian topological states have long been theorized, their (likely) realization has been limited to systems with strong magnetic field and low temperatures, such as LL systems \cite{Nayak2008}. Our findings extend the possibility of observing these states in a more favorable setting in moir\'e materials without the need for external fields, which has recently been utilized for realizing Abelian topological states at zero magnetic field and relatively high temperatures. From a theoretical standpoint, although the MR states represent the simplest manifestation of non-Abelian topological states, our work can be readily adapted to more complicate scenarios, such as the Read-Rezayi states at fractional fillings $2/5$ and $3/5$~\cite{read_rezayi}. 

Secondly, while evidence of charge density wave is observed both from entanglement spectrum and structure factor, the prominent peaks outside the MBZ suggest the potential existence of novel phases of matter. Finally, the transition between the CFL and MR states warrants further investigation.



\acknowledgments{{\it Acknowledgements.~---} We acknowledge useful discussions and related collaborations with Ahmed Abouelkomsan, Liang Fu, Aidan Reddy and Donna Sheng. H. Liu and E. J. Bergholtz were supported by the Swedish Research Council (VR, grant 2018-00313), the Wallenberg Academy Fellows program of the Knut and Alice Wallenberg Foundation (2018.0460) and the G\"oran Gustafsson Foundation for Research in Natural Sciences and Medicine. Z. Liu was supported by the National Natural Science Foundation of China (Grant No. 12350403 and 12374149).}

\bibliography{reference}

\begin{thebibliography}{83}%
\makeatletter
\providecommand \@ifxundefined [1]{%
 \@ifx{#1\undefined}
}%
\providecommand \@ifnum [1]{%
 \ifnum #1\expandafter \@firstoftwo
 \else \expandafter \@secondoftwo
 \fi
}%
\providecommand \@ifx [1]{%
 \ifx #1\expandafter \@firstoftwo
 \else \expandafter \@secondoftwo
 \fi
}%
\providecommand \natexlab [1]{#1}%
\providecommand \enquote  [1]{``#1''}%
\providecommand \bibnamefont  [1]{#1}%
\providecommand \bibfnamefont [1]{#1}%
\providecommand \citenamefont [1]{#1}%
\providecommand \href@noop [0]{\@secondoftwo}%
\providecommand \href [0]{\begingroup \@sanitize@url \@href}%
\providecommand \@href[1]{\@@startlink{#1}\@@href}%
\providecommand \@@href[1]{\endgroup#1\@@endlink}%
\providecommand \@sanitize@url [0]{\catcode `\\12\catcode `\$12\catcode `\&12\catcode `\#12\catcode `\^12\catcode `\_12\catcode `\%12\relax}%
\providecommand \@@startlink[1]{}%
\providecommand \@@endlink[0]{}%
\providecommand \url  [0]{\begingroup\@sanitize@url \@url }%
\providecommand \@url [1]{\endgroup\@href {#1}{\urlprefix }}%
\providecommand \urlprefix  [0]{URL }%
\providecommand \Eprint [0]{\href }%
\providecommand \doibase [0]{https://doi.org/}%
\providecommand \selectlanguage [0]{\@gobble}%
\providecommand \bibinfo  [0]{\@secondoftwo}%
\providecommand \bibfield  [0]{\@secondoftwo}%
\providecommand \translation [1]{[#1]}%
\providecommand \BibitemOpen [0]{}%
\providecommand \bibitemStop [0]{}%
\providecommand \bibitemNoStop [0]{.\EOS\space}%
\providecommand \EOS [0]{\spacefactor3000\relax}%
\providecommand \BibitemShut  [1]{\csname bibitem#1\endcsname}%
\let\auto@bib@innerbib\@empty
\bibitem [{\citenamefont {Andrei}\ and\ \citenamefont {MacDonald}(2020)}]{andrei2020graphene}%
  \BibitemOpen
  \bibfield  {author} {\bibinfo {author} {\bibfnamefont {E.~Y.}\ \bibnamefont {Andrei}}\ and\ \bibinfo {author} {\bibfnamefont {A.~H.}\ \bibnamefont {MacDonald}},\ }\bibfield  {title} {\bibinfo {title} {Graphene bilayers with a twist},\ }\href {https://doi.org/https://doi.org/10.1038/s41563-020-00840-0} {\bibfield  {journal} {\bibinfo  {journal} {Nature materials}\ }\textbf {\bibinfo {volume} {19}},\ \bibinfo {pages} {1265--1275} (\bibinfo {year} {2020})}\BibitemShut {NoStop}%
\bibitem [{\citenamefont {Andrei}\ \emph {et~al.}(2021)\citenamefont {Andrei}, \citenamefont {Efetov}, \citenamefont {Jarillo-Herrero}, \citenamefont {MacDonald}, \citenamefont {Mak}, \citenamefont {Senthil}, \citenamefont {Tutuc}, \citenamefont {Yazdani},\ and\ \citenamefont {Young}}]{Andrei_review}%
  \BibitemOpen
  \bibfield  {author} {\bibinfo {author} {\bibfnamefont {E.~Y.}\ \bibnamefont {Andrei}}, \bibinfo {author} {\bibfnamefont {D.~K.}\ \bibnamefont {Efetov}}, \bibinfo {author} {\bibfnamefont {P.}~\bibnamefont {Jarillo-Herrero}}, \bibinfo {author} {\bibfnamefont {A.~H.}\ \bibnamefont {MacDonald}}, \bibinfo {author} {\bibfnamefont {K.~F.}\ \bibnamefont {Mak}}, \bibinfo {author} {\bibfnamefont {T.}~\bibnamefont {Senthil}}, \bibinfo {author} {\bibfnamefont {E.}~\bibnamefont {Tutuc}}, \bibinfo {author} {\bibfnamefont {A.}~\bibnamefont {Yazdani}},\ and\ \bibinfo {author} {\bibfnamefont {A.~F.}\ \bibnamefont {Young}},\ }\bibfield  {title} {\bibinfo {title} {The marvels of moiré materials},\ }\href {https://doi.org/10.1038/s41578-021-00284-1} {\bibfield  {journal} {\bibinfo  {journal} {Nature Reviews Materials}\ }\textbf {\bibinfo {volume} {6}},\ \bibinfo {pages} {201--206} (\bibinfo {year} {2021})}\BibitemShut {NoStop}%
\bibitem [{\citenamefont {Cao}\ \emph {et~al.}(2018{\natexlab{a}})\citenamefont {Cao}, \citenamefont {Fatemi}, \citenamefont {Fang}, \citenamefont {Watanabe}, \citenamefont {Taniguchi}, \citenamefont {Kaxiras},\ and\ \citenamefont {{Jarillo-Herrero}}}]{caoUnconventionalSuperconductivityMagicangle2018}%
  \BibitemOpen
  \bibfield  {author} {\bibinfo {author} {\bibfnamefont {Y.}~\bibnamefont {Cao}}, \bibinfo {author} {\bibfnamefont {V.}~\bibnamefont {Fatemi}}, \bibinfo {author} {\bibfnamefont {S.}~\bibnamefont {Fang}}, \bibinfo {author} {\bibfnamefont {K.}~\bibnamefont {Watanabe}}, \bibinfo {author} {\bibfnamefont {T.}~\bibnamefont {Taniguchi}}, \bibinfo {author} {\bibfnamefont {E.}~\bibnamefont {Kaxiras}},\ and\ \bibinfo {author} {\bibfnamefont {P.}~\bibnamefont {{Jarillo-Herrero}}},\ }\bibfield  {title} {\bibinfo {title} {Unconventional superconductivity in magic-angle graphene superlattices},\ }\href {https://doi.org/10.1038/nature26160} {\bibfield  {journal} {\bibinfo  {journal} {Nature}\ }\textbf {\bibinfo {volume} {556}},\ \bibinfo {pages} {43--50} (\bibinfo {year} {2018}{\natexlab{a}})}\BibitemShut {NoStop}%
\bibitem [{\citenamefont {Cao}\ \emph {et~al.}(2018{\natexlab{b}})\citenamefont {Cao}, \citenamefont {Fatemi}, \citenamefont {Demir}, \citenamefont {Fang}, \citenamefont {Tomarken}, \citenamefont {Luo}, \citenamefont {Sanchez-Yamagishi}, \citenamefont {Watanabe}, \citenamefont {Taniguchi}, \citenamefont {Kaxiras}, \citenamefont {Ashoori},\ and\ \citenamefont {Jarillo-Herrero}}]{Cao2018}%
  \BibitemOpen
  \bibfield  {author} {\bibinfo {author} {\bibfnamefont {Y.}~\bibnamefont {Cao}}, \bibinfo {author} {\bibfnamefont {V.}~\bibnamefont {Fatemi}}, \bibinfo {author} {\bibfnamefont {A.}~\bibnamefont {Demir}}, \bibinfo {author} {\bibfnamefont {S.}~\bibnamefont {Fang}}, \bibinfo {author} {\bibfnamefont {S.~L.}\ \bibnamefont {Tomarken}}, \bibinfo {author} {\bibfnamefont {J.~Y.}\ \bibnamefont {Luo}}, \bibinfo {author} {\bibfnamefont {J.~D.}\ \bibnamefont {Sanchez-Yamagishi}}, \bibinfo {author} {\bibfnamefont {K.}~\bibnamefont {Watanabe}}, \bibinfo {author} {\bibfnamefont {T.}~\bibnamefont {Taniguchi}}, \bibinfo {author} {\bibfnamefont {E.}~\bibnamefont {Kaxiras}}, \bibinfo {author} {\bibfnamefont {R.~C.}\ \bibnamefont {Ashoori}},\ and\ \bibinfo {author} {\bibfnamefont {P.}~\bibnamefont {Jarillo-Herrero}},\ }\bibfield  {title} {\bibinfo {title} {Correlated insulator behaviour at half-filling in magic-angle graphene superlattices},\ }\href {https://doi.org/10.1038/nature26154} {\bibfield  {journal} {\bibinfo  {journal}
  {Nature}\ }\textbf {\bibinfo {volume} {556}},\ \bibinfo {pages} {80--84} (\bibinfo {year} {2018}{\natexlab{b}})}\BibitemShut {NoStop}%
\bibitem [{\citenamefont {Bistritzer}\ and\ \citenamefont {MacDonald}(2011)}]{Bistritzer2011}%
  \BibitemOpen
  \bibfield  {author} {\bibinfo {author} {\bibfnamefont {R.}~\bibnamefont {Bistritzer}}\ and\ \bibinfo {author} {\bibfnamefont {A.~H.}\ \bibnamefont {MacDonald}},\ }\bibfield  {title} {\bibinfo {title} {Moiré bands in twisted double-layer graphene},\ }\href {https://doi.org/10.1073/pnas.1108174108} {\bibfield  {journal} {\bibinfo  {journal} {Proceedings of the National Academy of Sciences}\ }\textbf {\bibinfo {volume} {108}},\ \bibinfo {pages} {12233--12237} (\bibinfo {year} {2011})},\ \Eprint {https://arxiv.org/abs/https://www.pnas.org/doi/pdf/10.1073/pnas.1108174108} {https://www.pnas.org/doi/pdf/10.1073/pnas.1108174108} \BibitemShut {NoStop}%
\bibitem [{\citenamefont {Liu}\ and\ \citenamefont {Bergholtz}(2024)}]{LIU2024515}%
  \BibitemOpen
  \bibfield  {author} {\bibinfo {author} {\bibfnamefont {Z.}~\bibnamefont {Liu}}\ and\ \bibinfo {author} {\bibfnamefont {E.~J.}\ \bibnamefont {Bergholtz}},\ }\bibfield  {title} {\bibinfo {title} {Recent developments in fractional {{Chern}} insulators},\ }in\ \href {https://doi.org/https://doi.org/10.1016/B978-0-323-90800-9.00136-0} {\emph {\bibinfo {booktitle} {Encyclopedia of Condensed Matter Physics (Second Edition)}}},\ \bibinfo {editor} {edited by\ \bibinfo {editor} {\bibfnamefont {T.}~\bibnamefont {Chakraborty}}}\ (\bibinfo  {publisher} {Academic Press},\ \bibinfo {address} {Oxford},\ \bibinfo {year} {2024})\ \bibinfo {edition} {second edition}\ ed.,\ pp.\ \bibinfo {pages} {515--538}\BibitemShut {NoStop}%
\bibitem [{\citenamefont {Abouelkomsan}\ \emph {et~al.}(2020)\citenamefont {Abouelkomsan}, \citenamefont {Liu},\ and\ \citenamefont {Bergholtz}}]{PhysRevLett.124.106803}%
  \BibitemOpen
  \bibfield  {author} {\bibinfo {author} {\bibfnamefont {A.}~\bibnamefont {Abouelkomsan}}, \bibinfo {author} {\bibfnamefont {Z.}~\bibnamefont {Liu}},\ and\ \bibinfo {author} {\bibfnamefont {E.~J.}\ \bibnamefont {Bergholtz}},\ }\bibfield  {title} {\bibinfo {title} {Particle-hole duality, emergent fermi liquids, and fractional {Chern} insulators in moir\'e flatbands},\ }\href {https://doi.org/10.1103/PhysRevLett.124.106803} {\bibfield  {journal} {\bibinfo  {journal} {Phys. Rev. Lett.}\ }\textbf {\bibinfo {volume} {124}},\ \bibinfo {pages} {106803} (\bibinfo {year} {2020})}\BibitemShut {NoStop}%
\bibitem [{\citenamefont {Repellin}\ and\ \citenamefont {Senthil}(2020)}]{repellinChernBandsTwisted2020}%
  \BibitemOpen
  \bibfield  {author} {\bibinfo {author} {\bibfnamefont {C.}~\bibnamefont {Repellin}}\ and\ \bibinfo {author} {\bibfnamefont {T.}~\bibnamefont {Senthil}},\ }\bibfield  {title} {\bibinfo {title} {Chern bands of twisted bilayer graphene: {{Fractional Chern}} insulators and spin phase transition},\ }\href {https://doi.org/10.1103/PhysRevResearch.2.023238} {\bibfield  {journal} {\bibinfo  {journal} {Physical Review Research}\ }\textbf {\bibinfo {volume} {2}},\ \bibinfo {pages} {023238} (\bibinfo {year} {2020})}\BibitemShut {NoStop}%
\bibitem [{\citenamefont {Ledwith}\ \emph {et~al.}(2020)\citenamefont {Ledwith}, \citenamefont {Tarnopolsky}, \citenamefont {Khalaf},\ and\ \citenamefont {Vishwanath}}]{PhysRevResearch.2.023237}%
  \BibitemOpen
  \bibfield  {author} {\bibinfo {author} {\bibfnamefont {P.~J.}\ \bibnamefont {Ledwith}}, \bibinfo {author} {\bibfnamefont {G.}~\bibnamefont {Tarnopolsky}}, \bibinfo {author} {\bibfnamefont {E.}~\bibnamefont {Khalaf}},\ and\ \bibinfo {author} {\bibfnamefont {A.}~\bibnamefont {Vishwanath}},\ }\bibfield  {title} {\bibinfo {title} {Fractional {Chern} insulator states in twisted bilayer graphene: An analytical approach},\ }\href {https://doi.org/10.1103/PhysRevResearch.2.023237} {\bibfield  {journal} {\bibinfo  {journal} {Phys. Rev. Research}\ }\textbf {\bibinfo {volume} {2}},\ \bibinfo {pages} {023237} (\bibinfo {year} {2020})}\BibitemShut {NoStop}%
\bibitem [{\citenamefont {Liu}\ \emph {et~al.}(2021)\citenamefont {Liu}, \citenamefont {Abouelkomsan},\ and\ \citenamefont {Bergholtz}}]{zhaoTDBG}%
  \BibitemOpen
  \bibfield  {author} {\bibinfo {author} {\bibfnamefont {Z.}~\bibnamefont {Liu}}, \bibinfo {author} {\bibfnamefont {A.}~\bibnamefont {Abouelkomsan}},\ and\ \bibinfo {author} {\bibfnamefont {E.~J.}\ \bibnamefont {Bergholtz}},\ }\bibfield  {title} {\bibinfo {title} {Gate-tunable fractional {{Chern}} insulators in twisted double bilayer graphene},\ }\href {https://doi.org/10.1103/PhysRevLett.126.026801} {\bibfield  {journal} {\bibinfo  {journal} {Phys. Rev. Lett.}\ }\textbf {\bibinfo {volume} {126}},\ \bibinfo {pages} {026801} (\bibinfo {year} {2021})}\BibitemShut {NoStop}%
\bibitem [{\citenamefont {Xie}\ \emph {et~al.}(2021)\citenamefont {Xie}, \citenamefont {Pierce}, \citenamefont {Park}, \citenamefont {Parker}, \citenamefont {Khalaf}, \citenamefont {Ledwith}, \citenamefont {Cao}, \citenamefont {Lee}, \citenamefont {Chen}, \citenamefont {Forrester} \emph {et~al.}}]{xie2021fractional}%
  \BibitemOpen
  \bibfield  {author} {\bibinfo {author} {\bibfnamefont {Y.}~\bibnamefont {Xie}}, \bibinfo {author} {\bibfnamefont {A.~T.}\ \bibnamefont {Pierce}}, \bibinfo {author} {\bibfnamefont {J.~M.}\ \bibnamefont {Park}}, \bibinfo {author} {\bibfnamefont {D.~E.}\ \bibnamefont {Parker}}, \bibinfo {author} {\bibfnamefont {E.}~\bibnamefont {Khalaf}}, \bibinfo {author} {\bibfnamefont {P.}~\bibnamefont {Ledwith}}, \bibinfo {author} {\bibfnamefont {Y.}~\bibnamefont {Cao}}, \bibinfo {author} {\bibfnamefont {S.~H.}\ \bibnamefont {Lee}}, \bibinfo {author} {\bibfnamefont {S.}~\bibnamefont {Chen}}, \bibinfo {author} {\bibfnamefont {P.~R.}\ \bibnamefont {Forrester}}, \emph {et~al.},\ }\bibfield  {title} {\bibinfo {title} {Fractional {Chern} insulators in magic-angle twisted bilayer graphene},\ }\href {https://doi.org/https://doi.org/10.1038/s41586-021-04002-3} {\bibfield  {journal} {\bibinfo  {journal} {Nature}\ }\textbf {\bibinfo {volume} {600}},\ \bibinfo {pages} {439--443} (\bibinfo {year} {2021})}\BibitemShut {NoStop}%
\bibitem [{\citenamefont {Lu}\ \emph {et~al.}(2024)\citenamefont {Lu}, \citenamefont {Han}, \citenamefont {Yao}, \citenamefont {Reddy}, \citenamefont {Yang}, \citenamefont {Seo}, \citenamefont {Watanabe}, \citenamefont {Taniguchi}, \citenamefont {Fu},\ and\ \citenamefont {Ju}}]{lu2023fractional}%
  \BibitemOpen
  \bibfield  {author} {\bibinfo {author} {\bibfnamefont {Z.}~\bibnamefont {Lu}}, \bibinfo {author} {\bibfnamefont {T.}~\bibnamefont {Han}}, \bibinfo {author} {\bibfnamefont {Y.}~\bibnamefont {Yao}}, \bibinfo {author} {\bibfnamefont {A.~P.}\ \bibnamefont {Reddy}}, \bibinfo {author} {\bibfnamefont {J.}~\bibnamefont {Yang}}, \bibinfo {author} {\bibfnamefont {J.}~\bibnamefont {Seo}}, \bibinfo {author} {\bibfnamefont {K.}~\bibnamefont {Watanabe}}, \bibinfo {author} {\bibfnamefont {T.}~\bibnamefont {Taniguchi}}, \bibinfo {author} {\bibfnamefont {L.}~\bibnamefont {Fu}},\ and\ \bibinfo {author} {\bibfnamefont {L.}~\bibnamefont {Ju}},\ }\bibfield  {title} {\bibinfo {title} {Fractional quantum anomalous hall effect in multilayer graphene},\ }\href {https://doi.org/10.1038/s41586-023-07010-7} {\bibfield  {journal} {\bibinfo  {journal} {Nature}\ }\textbf {\bibinfo {volume} {626}},\ \bibinfo {pages} {759--764} (\bibinfo {year} {2024})}\BibitemShut {NoStop}%
\bibitem [{\citenamefont {Spanton}\ \emph {et~al.}(2018)\citenamefont {Spanton}, \citenamefont {Zibrov}, \citenamefont {Zhou}, \citenamefont {Taniguchi}, \citenamefont {Watanabe}, \citenamefont {Zaletel},\ and\ \citenamefont {Young}}]{spanton2018observation}%
  \BibitemOpen
  \bibfield  {author} {\bibinfo {author} {\bibfnamefont {E.~M.}\ \bibnamefont {Spanton}}, \bibinfo {author} {\bibfnamefont {A.~A.}\ \bibnamefont {Zibrov}}, \bibinfo {author} {\bibfnamefont {H.}~\bibnamefont {Zhou}}, \bibinfo {author} {\bibfnamefont {T.}~\bibnamefont {Taniguchi}}, \bibinfo {author} {\bibfnamefont {K.}~\bibnamefont {Watanabe}}, \bibinfo {author} {\bibfnamefont {M.~P.}\ \bibnamefont {Zaletel}},\ and\ \bibinfo {author} {\bibfnamefont {A.~F.}\ \bibnamefont {Young}},\ }\bibfield  {title} {\bibinfo {title} {Observation of fractional {{Chern}} insulators in a van der waals heterostructure},\ }\href {https://www.science.org/doi/10.1126/science.aan8458} {\bibfield  {journal} {\bibinfo  {journal} {Science}\ }\textbf {\bibinfo {volume} {360}},\ \bibinfo {pages} {62--66} (\bibinfo {year} {2018})}\BibitemShut {NoStop}%
\bibitem [{\citenamefont {Li}\ \emph {et~al.}(2021)\citenamefont {Li}, \citenamefont {Kumar}, \citenamefont {Sun},\ and\ \citenamefont {Lin}}]{PhysRevResearch.3.L032070}%
  \BibitemOpen
  \bibfield  {author} {\bibinfo {author} {\bibfnamefont {H.}~\bibnamefont {Li}}, \bibinfo {author} {\bibfnamefont {U.}~\bibnamefont {Kumar}}, \bibinfo {author} {\bibfnamefont {K.}~\bibnamefont {Sun}},\ and\ \bibinfo {author} {\bibfnamefont {S.-Z.}\ \bibnamefont {Lin}},\ }\bibfield  {title} {\bibinfo {title} {Spontaneous fractional {{Chern}} insulators in transition metal dichalcogenide moir\'e superlattices},\ }\href {https://doi.org/10.1103/PhysRevResearch.3.L032070} {\bibfield  {journal} {\bibinfo  {journal} {Phys. Rev. Res.}\ }\textbf {\bibinfo {volume} {3}},\ \bibinfo {pages} {L032070} (\bibinfo {year} {2021})}\BibitemShut {NoStop}%
\bibitem [{\citenamefont {Cr\'epel}\ and\ \citenamefont {Fu}(2023)}]{PhysRevB.107.L201109}%
  \BibitemOpen
  \bibfield  {author} {\bibinfo {author} {\bibfnamefont {V.}~\bibnamefont {Cr\'epel}}\ and\ \bibinfo {author} {\bibfnamefont {L.}~\bibnamefont {Fu}},\ }\bibfield  {title} {\bibinfo {title} {Anomalous hall metal and fractional {{Chern}} insulator in twisted transition metal dichalcogenides},\ }\href {https://doi.org/10.1103/PhysRevB.107.L201109} {\bibfield  {journal} {\bibinfo  {journal} {Phys. Rev. B}\ }\textbf {\bibinfo {volume} {107}},\ \bibinfo {pages} {L201109} (\bibinfo {year} {2023})}\BibitemShut {NoStop}%
\bibitem [{\citenamefont {Cai}\ \emph {et~al.}(2023)\citenamefont {Cai}, \citenamefont {Anderson}, \citenamefont {Wang}, \citenamefont {Zhang}, \citenamefont {Liu}, \citenamefont {Holtzmann}, \citenamefont {Zhang}, \citenamefont {Fan}, \citenamefont {Taniguchi}, \citenamefont {Watanabe}, \citenamefont {Ran}, \citenamefont {Cao}, \citenamefont {Fu}, \citenamefont {Xiao}, \citenamefont {Yao},\ and\ \citenamefont {Xu}}]{FCI_MoTe2_1}%
  \BibitemOpen
  \bibfield  {author} {\bibinfo {author} {\bibfnamefont {J.}~\bibnamefont {Cai}}, \bibinfo {author} {\bibfnamefont {E.}~\bibnamefont {Anderson}}, \bibinfo {author} {\bibfnamefont {C.}~\bibnamefont {Wang}}, \bibinfo {author} {\bibfnamefont {X.}~\bibnamefont {Zhang}}, \bibinfo {author} {\bibfnamefont {X.}~\bibnamefont {Liu}}, \bibinfo {author} {\bibfnamefont {W.}~\bibnamefont {Holtzmann}}, \bibinfo {author} {\bibfnamefont {Y.}~\bibnamefont {Zhang}}, \bibinfo {author} {\bibfnamefont {F.}~\bibnamefont {Fan}}, \bibinfo {author} {\bibfnamefont {T.}~\bibnamefont {Taniguchi}}, \bibinfo {author} {\bibfnamefont {K.}~\bibnamefont {Watanabe}}, \bibinfo {author} {\bibfnamefont {Y.}~\bibnamefont {Ran}}, \bibinfo {author} {\bibfnamefont {T.}~\bibnamefont {Cao}}, \bibinfo {author} {\bibfnamefont {L.}~\bibnamefont {Fu}}, \bibinfo {author} {\bibfnamefont {D.}~\bibnamefont {Xiao}}, \bibinfo {author} {\bibfnamefont {W.}~\bibnamefont {Yao}},\ and\ \bibinfo {author} {\bibfnamefont {X.}~\bibnamefont {Xu}},\ }\bibfield  {title}
  {\bibinfo {title} {Signatures of fractional quantum anomalous hall states in twisted mote2},\ }\href {https://doi.org/10.1038/s41586-023-06289-w} {\bibfield  {journal} {\bibinfo  {journal} {Nature}\ }\textbf {\bibinfo {volume} {622}},\ \bibinfo {pages} {63--68} (\bibinfo {year} {2023})}\BibitemShut {NoStop}%
\bibitem [{\citenamefont {Zeng}\ \emph {et~al.}(2023)\citenamefont {Zeng}, \citenamefont {Xia}, \citenamefont {Kang}, \citenamefont {Zhu}, \citenamefont {Kn{\"u}ppel}, \citenamefont {Vaswani}, \citenamefont {Watanabe}, \citenamefont {Taniguchi}, \citenamefont {Mak},\ and\ \citenamefont {Shan}}]{FCI_MoTe2_2}%
  \BibitemOpen
  \bibfield  {author} {\bibinfo {author} {\bibfnamefont {Y.}~\bibnamefont {Zeng}}, \bibinfo {author} {\bibfnamefont {Z.}~\bibnamefont {Xia}}, \bibinfo {author} {\bibfnamefont {K.}~\bibnamefont {Kang}}, \bibinfo {author} {\bibfnamefont {J.}~\bibnamefont {Zhu}}, \bibinfo {author} {\bibfnamefont {P.}~\bibnamefont {Kn{\"u}ppel}}, \bibinfo {author} {\bibfnamefont {C.}~\bibnamefont {Vaswani}}, \bibinfo {author} {\bibfnamefont {K.}~\bibnamefont {Watanabe}}, \bibinfo {author} {\bibfnamefont {T.}~\bibnamefont {Taniguchi}}, \bibinfo {author} {\bibfnamefont {K.~F.}\ \bibnamefont {Mak}},\ and\ \bibinfo {author} {\bibfnamefont {J.}~\bibnamefont {Shan}},\ }\bibfield  {title} {\bibinfo {title} {Thermodynamic evidence of fractional {Chern} insulator in moir{\'e}mote2},\ }\href {https://doi.org/10.1038/s41586-023-06452-3} {\bibfield  {journal} {\bibinfo  {journal} {Nature}\ }\textbf {\bibinfo {volume} {622}},\ \bibinfo {pages} {69--73} (\bibinfo {year} {2023})}\BibitemShut {NoStop}%
\bibitem [{\citenamefont {Park}\ \emph {et~al.}(2023)\citenamefont {Park}, \citenamefont {Cai}, \citenamefont {Anderson}, \citenamefont {Zhang}, \citenamefont {Zhu}, \citenamefont {Liu}, \citenamefont {Wang}, \citenamefont {Holtzmann}, \citenamefont {Hu}, \citenamefont {Liu}, \citenamefont {Taniguchi}, \citenamefont {Watanabe}, \citenamefont {Chu}, \citenamefont {Cao}, \citenamefont {Fu}, \citenamefont {Yao}, \citenamefont {Chang}, \citenamefont {Cobden}, \citenamefont {Xiao},\ and\ \citenamefont {Xu}}]{FCI_MoTe2_3}%
  \BibitemOpen
  \bibfield  {author} {\bibinfo {author} {\bibfnamefont {H.}~\bibnamefont {Park}}, \bibinfo {author} {\bibfnamefont {J.}~\bibnamefont {Cai}}, \bibinfo {author} {\bibfnamefont {E.}~\bibnamefont {Anderson}}, \bibinfo {author} {\bibfnamefont {Y.}~\bibnamefont {Zhang}}, \bibinfo {author} {\bibfnamefont {J.}~\bibnamefont {Zhu}}, \bibinfo {author} {\bibfnamefont {X.}~\bibnamefont {Liu}}, \bibinfo {author} {\bibfnamefont {C.}~\bibnamefont {Wang}}, \bibinfo {author} {\bibfnamefont {W.}~\bibnamefont {Holtzmann}}, \bibinfo {author} {\bibfnamefont {C.}~\bibnamefont {Hu}}, \bibinfo {author} {\bibfnamefont {Z.}~\bibnamefont {Liu}}, \bibinfo {author} {\bibfnamefont {T.}~\bibnamefont {Taniguchi}}, \bibinfo {author} {\bibfnamefont {K.}~\bibnamefont {Watanabe}}, \bibinfo {author} {\bibfnamefont {J.-H.}\ \bibnamefont {Chu}}, \bibinfo {author} {\bibfnamefont {T.}~\bibnamefont {Cao}}, \bibinfo {author} {\bibfnamefont {L.}~\bibnamefont {Fu}}, \bibinfo {author} {\bibfnamefont {W.}~\bibnamefont {Yao}}, \bibinfo {author}
  {\bibfnamefont {C.-Z.}\ \bibnamefont {Chang}}, \bibinfo {author} {\bibfnamefont {D.}~\bibnamefont {Cobden}}, \bibinfo {author} {\bibfnamefont {D.}~\bibnamefont {Xiao}},\ and\ \bibinfo {author} {\bibfnamefont {X.}~\bibnamefont {Xu}},\ }\bibfield  {title} {\bibinfo {title} {Observation of fractionally quantized anomalous hall effect},\ }\href {https://doi.org/10.1038/s41586-023-06536-0} {\bibfield  {journal} {\bibinfo  {journal} {Nature}\ }\textbf {\bibinfo {volume} {622}},\ \bibinfo {pages} {74--79} (\bibinfo {year} {2023})}\BibitemShut {NoStop}%
\bibitem [{\citenamefont {Xu}\ \emph {et~al.}(2023)\citenamefont {Xu}, \citenamefont {Sun}, \citenamefont {Jia}, \citenamefont {Liu}, \citenamefont {Xu}, \citenamefont {Li}, \citenamefont {Gu}, \citenamefont {Watanabe}, \citenamefont {Taniguchi}, \citenamefont {Tong}, \citenamefont {Jia}, \citenamefont {Shi}, \citenamefont {Jiang}, \citenamefont {Zhang}, \citenamefont {Liu},\ and\ \citenamefont {Li}}]{PhysRevX.13.031037}%
  \BibitemOpen
  \bibfield  {author} {\bibinfo {author} {\bibfnamefont {F.}~\bibnamefont {Xu}}, \bibinfo {author} {\bibfnamefont {Z.}~\bibnamefont {Sun}}, \bibinfo {author} {\bibfnamefont {T.}~\bibnamefont {Jia}}, \bibinfo {author} {\bibfnamefont {C.}~\bibnamefont {Liu}}, \bibinfo {author} {\bibfnamefont {C.}~\bibnamefont {Xu}}, \bibinfo {author} {\bibfnamefont {C.}~\bibnamefont {Li}}, \bibinfo {author} {\bibfnamefont {Y.}~\bibnamefont {Gu}}, \bibinfo {author} {\bibfnamefont {K.}~\bibnamefont {Watanabe}}, \bibinfo {author} {\bibfnamefont {T.}~\bibnamefont {Taniguchi}}, \bibinfo {author} {\bibfnamefont {B.}~\bibnamefont {Tong}}, \bibinfo {author} {\bibfnamefont {J.}~\bibnamefont {Jia}}, \bibinfo {author} {\bibfnamefont {Z.}~\bibnamefont {Shi}}, \bibinfo {author} {\bibfnamefont {S.}~\bibnamefont {Jiang}}, \bibinfo {author} {\bibfnamefont {Y.}~\bibnamefont {Zhang}}, \bibinfo {author} {\bibfnamefont {X.}~\bibnamefont {Liu}},\ and\ \bibinfo {author} {\bibfnamefont {T.}~\bibnamefont {Li}},\ }\bibfield  {title} {\bibinfo {title}
  {Observation of integer and fractional quantum anomalous {Hall} effects in twisted bilayer ${\mathrm{mote}}_{2}$},\ }\href {https://doi.org/10.1103/PhysRevX.13.031037} {\bibfield  {journal} {\bibinfo  {journal} {Phys. Rev. X}\ }\textbf {\bibinfo {volume} {13}},\ \bibinfo {pages} {031037} (\bibinfo {year} {2023})}\BibitemShut {NoStop}%
\bibitem [{\citenamefont {Nayak}\ \emph {et~al.}(2008)\citenamefont {Nayak}, \citenamefont {Simon}, \citenamefont {Stern}, \citenamefont {Freedman},\ and\ \citenamefont {Das~Sarma}}]{Nayak2008}%
  \BibitemOpen
  \bibfield  {author} {\bibinfo {author} {\bibfnamefont {C.}~\bibnamefont {Nayak}}, \bibinfo {author} {\bibfnamefont {S.~H.}\ \bibnamefont {Simon}}, \bibinfo {author} {\bibfnamefont {A.}~\bibnamefont {Stern}}, \bibinfo {author} {\bibfnamefont {M.}~\bibnamefont {Freedman}},\ and\ \bibinfo {author} {\bibfnamefont {S.}~\bibnamefont {Das~Sarma}},\ }\bibfield  {title} {\bibinfo {title} {Non-abelian anyons and topological quantum computation},\ }\href {https://doi.org/10.1103/RevModPhys.80.1083} {\bibfield  {journal} {\bibinfo  {journal} {Rev. Mod. Phys.}\ }\textbf {\bibinfo {volume} {80}},\ \bibinfo {pages} {1083--1159} (\bibinfo {year} {2008})}\BibitemShut {NoStop}%
\bibitem [{\citenamefont {Bergholtz}\ and\ \citenamefont {Liu}(2013)}]{bergholtz2013topological}%
  \BibitemOpen
  \bibfield  {author} {\bibinfo {author} {\bibfnamefont {E.~J.}\ \bibnamefont {Bergholtz}}\ and\ \bibinfo {author} {\bibfnamefont {Z.}~\bibnamefont {Liu}},\ }\bibfield  {title} {\bibinfo {title} {Topological flat band models and fractional chern insulators},\ }\href {https://doi.org/https://doi.org/10.1142/S021797921330017X} {\bibfield  {journal} {\bibinfo  {journal} {International Journal of Modern Physics B}\ }\textbf {\bibinfo {volume} {27}},\ \bibinfo {pages} {1330017} (\bibinfo {year} {2013})}\BibitemShut {NoStop}%
\bibitem [{\citenamefont {Parameswaran}\ \emph {et~al.}(2013)\citenamefont {Parameswaran}, \citenamefont {Roy},\ and\ \citenamefont {Sondhi}}]{PARAMESWARAN2013816}%
  \BibitemOpen
  \bibfield  {author} {\bibinfo {author} {\bibfnamefont {S.~A.}\ \bibnamefont {Parameswaran}}, \bibinfo {author} {\bibfnamefont {R.}~\bibnamefont {Roy}},\ and\ \bibinfo {author} {\bibfnamefont {S.~L.}\ \bibnamefont {Sondhi}},\ }\bibfield  {title} {\bibinfo {title} {Fractional quantum hall physics in topological flat bands},\ }\href {https://doi.org/https://doi.org/10.1016/j.crhy.2013.04.003} {\bibfield  {journal} {\bibinfo  {journal} {Comptes Rendus Physique}\ }\textbf {\bibinfo {volume} {14}},\ \bibinfo {pages} {816--839} (\bibinfo {year} {2013})},\ \bibinfo {note} {topological insulators / Isolants topologiques}\BibitemShut {NoStop}%
\bibitem [{\citenamefont {Kol}\ and\ \citenamefont {Read}(1993)}]{kolread}%
  \BibitemOpen
  \bibfield  {author} {\bibinfo {author} {\bibfnamefont {A.}~\bibnamefont {Kol}}\ and\ \bibinfo {author} {\bibfnamefont {N.}~\bibnamefont {Read}},\ }\bibfield  {title} {\bibinfo {title} {Fractional quantum hall effect in a periodic potential},\ }\href {https://doi.org/10.1103/PhysRevB.48.8890} {\bibfield  {journal} {\bibinfo  {journal} {Phys. Rev. B}\ }\textbf {\bibinfo {volume} {48}},\ \bibinfo {pages} {8890--8898} (\bibinfo {year} {1993})}\BibitemShut {NoStop}%
\bibitem [{\citenamefont {Tang}\ \emph {et~al.}(2011)\citenamefont {Tang}, \citenamefont {Mei},\ and\ \citenamefont {Wen}}]{tang_high-temperature_2011}%
  \BibitemOpen
  \bibfield  {author} {\bibinfo {author} {\bibfnamefont {E.}~\bibnamefont {Tang}}, \bibinfo {author} {\bibfnamefont {J.-W.}\ \bibnamefont {Mei}},\ and\ \bibinfo {author} {\bibfnamefont {X.-G.}\ \bibnamefont {Wen}},\ }\bibfield  {title} {\bibinfo {title} {High-{Temperature} {Fractional} {Quantum} {Hall} {States}},\ }\href {https://doi.org/10.1103/PhysRevLett.106.236802} {\bibfield  {journal} {\bibinfo  {journal} {Physical Review Letters}\ }\textbf {\bibinfo {volume} {106}},\ \bibinfo {pages} {236802} (\bibinfo {year} {2011})}\BibitemShut {NoStop}%
\bibitem [{\citenamefont {Sun}\ \emph {et~al.}(2011)\citenamefont {Sun}, \citenamefont {Gu}, \citenamefont {Katsura},\ and\ \citenamefont {Das~Sarma}}]{PhysRevLett.106.236803}%
  \BibitemOpen
  \bibfield  {author} {\bibinfo {author} {\bibfnamefont {K.}~\bibnamefont {Sun}}, \bibinfo {author} {\bibfnamefont {Z.}~\bibnamefont {Gu}}, \bibinfo {author} {\bibfnamefont {H.}~\bibnamefont {Katsura}},\ and\ \bibinfo {author} {\bibfnamefont {S.}~\bibnamefont {Das~Sarma}},\ }\bibfield  {title} {\bibinfo {title} {Nearly flatbands with nontrivial topology},\ }\href {https://doi.org/10.1103/PhysRevLett.106.236803} {\bibfield  {journal} {\bibinfo  {journal} {Phys. Rev. Lett.}\ }\textbf {\bibinfo {volume} {106}},\ \bibinfo {pages} {236803} (\bibinfo {year} {2011})}\BibitemShut {NoStop}%
\bibitem [{\citenamefont {Neupert}\ \emph {et~al.}(2011)\citenamefont {Neupert}, \citenamefont {Santos}, \citenamefont {Chamon},\ and\ \citenamefont {Mudry}}]{PhysRevLett.106.236804}%
  \BibitemOpen
  \bibfield  {author} {\bibinfo {author} {\bibfnamefont {T.}~\bibnamefont {Neupert}}, \bibinfo {author} {\bibfnamefont {L.}~\bibnamefont {Santos}}, \bibinfo {author} {\bibfnamefont {C.}~\bibnamefont {Chamon}},\ and\ \bibinfo {author} {\bibfnamefont {C.}~\bibnamefont {Mudry}},\ }\bibfield  {title} {\bibinfo {title} {Fractional quantum {Hall} states at zero magnetic field},\ }\href {https://doi.org/10.1103/PhysRevLett.106.236804} {\bibfield  {journal} {\bibinfo  {journal} {Phys. Rev. Lett.}\ }\textbf {\bibinfo {volume} {106}},\ \bibinfo {pages} {236804} (\bibinfo {year} {2011})}\BibitemShut {NoStop}%
\bibitem [{\citenamefont {Sheng}\ \emph {et~al.}(2011)\citenamefont {Sheng}, \citenamefont {Gu}, \citenamefont {Sun},\ and\ \citenamefont {Sheng}}]{sheng2011fractional}%
  \BibitemOpen
  \bibfield  {author} {\bibinfo {author} {\bibfnamefont {D.}~\bibnamefont {Sheng}}, \bibinfo {author} {\bibfnamefont {Z.-C.}\ \bibnamefont {Gu}}, \bibinfo {author} {\bibfnamefont {K.}~\bibnamefont {Sun}},\ and\ \bibinfo {author} {\bibfnamefont {L.}~\bibnamefont {Sheng}},\ }\bibfield  {title} {\bibinfo {title} {Fractional quantum {Hall} effect in the absence of landau levels},\ }\href {https://doi.org/https://doi.org/10.1038/ncomms1380} {\bibfield  {journal} {\bibinfo  {journal} {Nature communications}\ }\textbf {\bibinfo {volume} {2}},\ \bibinfo {pages} {1--5} (\bibinfo {year} {2011})}\BibitemShut {NoStop}%
\bibitem [{\citenamefont {Regnault}\ and\ \citenamefont {Bernevig}(2011)}]{PhysRevX.1.021014}%
  \BibitemOpen
  \bibfield  {author} {\bibinfo {author} {\bibfnamefont {N.}~\bibnamefont {Regnault}}\ and\ \bibinfo {author} {\bibfnamefont {B.~A.}\ \bibnamefont {Bernevig}},\ }\bibfield  {title} {\bibinfo {title} {Fractional chern insulator},\ }\href {https://doi.org/10.1103/PhysRevX.1.021014} {\bibfield  {journal} {\bibinfo  {journal} {Phys. Rev. X}\ }\textbf {\bibinfo {volume} {1}},\ \bibinfo {pages} {021014} (\bibinfo {year} {2011})}\BibitemShut {NoStop}%
\bibitem [{\citenamefont {M\"oller}\ and\ \citenamefont {Cooper}(2009)}]{mollercooper}%
  \BibitemOpen
  \bibfield  {author} {\bibinfo {author} {\bibfnamefont {G.}~\bibnamefont {M\"oller}}\ and\ \bibinfo {author} {\bibfnamefont {N.~R.}\ \bibnamefont {Cooper}},\ }\bibfield  {title} {\bibinfo {title} {Composite fermion theory for bosonic quantum hall states on lattices},\ }\href {https://doi.org/10.1103/PhysRevLett.103.105303} {\bibfield  {journal} {\bibinfo  {journal} {Phys. Rev. Lett.}\ }\textbf {\bibinfo {volume} {103}},\ \bibinfo {pages} {105303} (\bibinfo {year} {2009})}\BibitemShut {NoStop}%
\bibitem [{\citenamefont {Kapit}\ and\ \citenamefont {Mueller}(2010)}]{PhysRevLett.105.215303}%
  \BibitemOpen
  \bibfield  {author} {\bibinfo {author} {\bibfnamefont {E.}~\bibnamefont {Kapit}}\ and\ \bibinfo {author} {\bibfnamefont {E.}~\bibnamefont {Mueller}},\ }\bibfield  {title} {\bibinfo {title} {Exact parent hamiltonian for the quantum hall states in a lattice},\ }\href {https://doi.org/10.1103/PhysRevLett.105.215303} {\bibfield  {journal} {\bibinfo  {journal} {Phys. Rev. Lett.}\ }\textbf {\bibinfo {volume} {105}},\ \bibinfo {pages} {215303} (\bibinfo {year} {2010})}\BibitemShut {NoStop}%
\bibitem [{\citenamefont {Wang}\ \emph {et~al.}(2012)\citenamefont {Wang}, \citenamefont {Yao}, \citenamefont {Gu}, \citenamefont {Gong},\ and\ \citenamefont {Sheng}}]{Wang2012}%
  \BibitemOpen
  \bibfield  {author} {\bibinfo {author} {\bibfnamefont {Y.-F.}\ \bibnamefont {Wang}}, \bibinfo {author} {\bibfnamefont {H.}~\bibnamefont {Yao}}, \bibinfo {author} {\bibfnamefont {Z.-C.}\ \bibnamefont {Gu}}, \bibinfo {author} {\bibfnamefont {C.-D.}\ \bibnamefont {Gong}},\ and\ \bibinfo {author} {\bibfnamefont {D.~N.}\ \bibnamefont {Sheng}},\ }\bibfield  {title} {\bibinfo {title} {Non-abelian quantum hall effect in topological flat bands},\ }\href {https://doi.org/10.1103/PhysRevLett.108.126805} {\bibfield  {journal} {\bibinfo  {journal} {Phys. Rev. Lett.}\ }\textbf {\bibinfo {volume} {108}},\ \bibinfo {pages} {126805} (\bibinfo {year} {2012})}\BibitemShut {NoStop}%
\bibitem [{\citenamefont {Wu}\ \emph {et~al.}(2012)\citenamefont {Wu}, \citenamefont {Bernevig},\ and\ \citenamefont {Regnault}}]{Zoology2012}%
  \BibitemOpen
  \bibfield  {author} {\bibinfo {author} {\bibfnamefont {Y.-L.}\ \bibnamefont {Wu}}, \bibinfo {author} {\bibfnamefont {B.~A.}\ \bibnamefont {Bernevig}},\ and\ \bibinfo {author} {\bibfnamefont {N.}~\bibnamefont {Regnault}},\ }\bibfield  {title} {\bibinfo {title} {Zoology of fractional chern insulators},\ }\href {https://doi.org/10.1103/PhysRevB.85.075116} {\bibfield  {journal} {\bibinfo  {journal} {Phys. Rev. B}\ }\textbf {\bibinfo {volume} {85}},\ \bibinfo {pages} {075116} (\bibinfo {year} {2012})}\BibitemShut {NoStop}%
\bibitem [{\citenamefont {Sterdyniak}\ \emph {et~al.}(2013)\citenamefont {Sterdyniak}, \citenamefont {Repellin}, \citenamefont {Bernevig},\ and\ \citenamefont {Regnault}}]{Repellin2013}%
  \BibitemOpen
  \bibfield  {author} {\bibinfo {author} {\bibfnamefont {A.}~\bibnamefont {Sterdyniak}}, \bibinfo {author} {\bibfnamefont {C.}~\bibnamefont {Repellin}}, \bibinfo {author} {\bibfnamefont {B.~A.}\ \bibnamefont {Bernevig}},\ and\ \bibinfo {author} {\bibfnamefont {N.}~\bibnamefont {Regnault}},\ }\bibfield  {title} {\bibinfo {title} {Series of abelian and non-abelian states in $c>1$ fractional chern insulators},\ }\href {https://doi.org/10.1103/PhysRevB.87.205137} {\bibfield  {journal} {\bibinfo  {journal} {Phys. Rev. B}\ }\textbf {\bibinfo {volume} {87}},\ \bibinfo {pages} {205137} (\bibinfo {year} {2013})}\BibitemShut {NoStop}%
\bibitem [{\citenamefont {Wu}\ \emph {et~al.}(2013)\citenamefont {Wu}, \citenamefont {Regnault},\ and\ \citenamefont {Bernevig}}]{Wu2013}%
  \BibitemOpen
  \bibfield  {author} {\bibinfo {author} {\bibfnamefont {Y.-L.}\ \bibnamefont {Wu}}, \bibinfo {author} {\bibfnamefont {N.}~\bibnamefont {Regnault}},\ and\ \bibinfo {author} {\bibfnamefont {B.~A.}\ \bibnamefont {Bernevig}},\ }\bibfield  {title} {\bibinfo {title} {Bloch model wave functions and pseudopotentials for all fractional chern insulators},\ }\href {https://doi.org/10.1103/PhysRevLett.110.106802} {\bibfield  {journal} {\bibinfo  {journal} {Phys. Rev. Lett.}\ }\textbf {\bibinfo {volume} {110}},\ \bibinfo {pages} {106802} (\bibinfo {year} {2013})}\BibitemShut {NoStop}%
\bibitem [{\citenamefont {Bergholtz}\ \emph {et~al.}(2015)\citenamefont {Bergholtz}, \citenamefont {Liu}, \citenamefont {Trescher}, \citenamefont {Moessner},\ and\ \citenamefont {Udagawa}}]{Bergholtz2015}%
  \BibitemOpen
  \bibfield  {author} {\bibinfo {author} {\bibfnamefont {E.~J.}\ \bibnamefont {Bergholtz}}, \bibinfo {author} {\bibfnamefont {Z.}~\bibnamefont {Liu}}, \bibinfo {author} {\bibfnamefont {M.}~\bibnamefont {Trescher}}, \bibinfo {author} {\bibfnamefont {R.}~\bibnamefont {Moessner}},\ and\ \bibinfo {author} {\bibfnamefont {M.}~\bibnamefont {Udagawa}},\ }\bibfield  {title} {\bibinfo {title} {Topology and interactions in a frustrated slab: Tuning from weyl semimetals to $\mathcal{C}>1$ fractional chern insulators},\ }\href {https://doi.org/10.1103/PhysRevLett.114.016806} {\bibfield  {journal} {\bibinfo  {journal} {Phys. Rev. Lett.}\ }\textbf {\bibinfo {volume} {114}},\ \bibinfo {pages} {016806} (\bibinfo {year} {2015})}\BibitemShut {NoStop}%
\bibitem [{\citenamefont {Behrmann}\ \emph {et~al.}(2016)\citenamefont {Behrmann}, \citenamefont {Liu},\ and\ \citenamefont {Bergholtz}}]{Behrmann2016}%
  \BibitemOpen
  \bibfield  {author} {\bibinfo {author} {\bibfnamefont {J.}~\bibnamefont {Behrmann}}, \bibinfo {author} {\bibfnamefont {Z.}~\bibnamefont {Liu}},\ and\ \bibinfo {author} {\bibfnamefont {E.~J.}\ \bibnamefont {Bergholtz}},\ }\bibfield  {title} {\bibinfo {title} {Model fractional chern insulators},\ }\href {https://doi.org/10.1103/PhysRevLett.116.216802} {\bibfield  {journal} {\bibinfo  {journal} {Phys. Rev. Lett.}\ }\textbf {\bibinfo {volume} {116}},\ \bibinfo {pages} {216802} (\bibinfo {year} {2016})}\BibitemShut {NoStop}%
\bibitem [{\citenamefont {Moore}\ and\ \citenamefont {Read}(1991)}]{MOORE1991362}%
  \BibitemOpen
  \bibfield  {author} {\bibinfo {author} {\bibfnamefont {G.}~\bibnamefont {Moore}}\ and\ \bibinfo {author} {\bibfnamefont {N.}~\bibnamefont {Read}},\ }\bibfield  {title} {\bibinfo {title} {Nonabelions in the fractional quantum hall effect},\ }\href {https://doi.org/https://doi.org/10.1016/0550-3213(91)90407-O} {\bibfield  {journal} {\bibinfo  {journal} {Nuclear Physics B}\ }\textbf {\bibinfo {volume} {360}},\ \bibinfo {pages} {362--396} (\bibinfo {year} {1991})}\BibitemShut {NoStop}%
\bibitem [{\citenamefont {Greiter}\ \emph {et~al.}(1992)\citenamefont {Greiter}, \citenamefont {Wen},\ and\ \citenamefont {Wilczek}}]{greiter1992}%
  \BibitemOpen
  \bibfield  {author} {\bibinfo {author} {\bibfnamefont {M.}~\bibnamefont {Greiter}}, \bibinfo {author} {\bibfnamefont {X.-G.}\ \bibnamefont {Wen}},\ and\ \bibinfo {author} {\bibfnamefont {F.}~\bibnamefont {Wilczek}},\ }\bibfield  {title} {\bibinfo {title} {Paired hall states},\ }\href {https://www.sciencedirect.com/science/article/pii/055032139290401V} {\bibfield  {journal} {\bibinfo  {journal} {Nuclear Physics B}\ }\textbf {\bibinfo {volume} {374}},\ \bibinfo {pages} {567--614} (\bibinfo {year} {1992})}\BibitemShut {NoStop}%
\bibitem [{\citenamefont {Read}\ and\ \citenamefont {Rezayi}(1999)}]{read_rezayi}%
  \BibitemOpen
  \bibfield  {author} {\bibinfo {author} {\bibfnamefont {N.}~\bibnamefont {Read}}\ and\ \bibinfo {author} {\bibfnamefont {E.}~\bibnamefont {Rezayi}},\ }\bibfield  {title} {\bibinfo {title} {Beyond paired quantum hall states: Parafermions and incompressible states in the first excited landau level},\ }\href {https://doi.org/10.1103/PhysRevB.59.8084} {\bibfield  {journal} {\bibinfo  {journal} {Phys. Rev. B}\ }\textbf {\bibinfo {volume} {59}},\ \bibinfo {pages} {8084--8092} (\bibinfo {year} {1999})}\BibitemShut {NoStop}%
\bibitem [{\citenamefont {Liu}\ \emph {et~al.}(2013)\citenamefont {Liu}, \citenamefont {Bergholtz},\ and\ \citenamefont {Kapit}}]{NAlong2013}%
  \BibitemOpen
  \bibfield  {author} {\bibinfo {author} {\bibfnamefont {Z.}~\bibnamefont {Liu}}, \bibinfo {author} {\bibfnamefont {E.~J.}\ \bibnamefont {Bergholtz}},\ and\ \bibinfo {author} {\bibfnamefont {E.}~\bibnamefont {Kapit}},\ }\bibfield  {title} {\bibinfo {title} {Non-abelian fractional {{Chern}} insulators from long-range interactions},\ }\href {https://doi.org/10.1103/PhysRevB.88.205101} {\bibfield  {journal} {\bibinfo  {journal} {Phys. Rev. B}\ }\textbf {\bibinfo {volume} {88}},\ \bibinfo {pages} {205101} (\bibinfo {year} {2013})}\BibitemShut {NoStop}%
\bibitem [{\citenamefont {Wang}\ \emph {et~al.}(2015)\citenamefont {Wang}, \citenamefont {Liu}, \citenamefont {Liu}, \citenamefont {Cao},\ and\ \citenamefont {Fan}}]{Wang2015}%
  \BibitemOpen
  \bibfield  {author} {\bibinfo {author} {\bibfnamefont {D.}~\bibnamefont {Wang}}, \bibinfo {author} {\bibfnamefont {Z.}~\bibnamefont {Liu}}, \bibinfo {author} {\bibfnamefont {W.-M.}\ \bibnamefont {Liu}}, \bibinfo {author} {\bibfnamefont {J.}~\bibnamefont {Cao}},\ and\ \bibinfo {author} {\bibfnamefont {H.}~\bibnamefont {Fan}},\ }\bibfield  {title} {\bibinfo {title} {Fermionic non-abelian fractional chern insulators from dipolar interactions},\ }\href {https://doi.org/10.1103/PhysRevB.91.125138} {\bibfield  {journal} {\bibinfo  {journal} {Phys. Rev. B}\ }\textbf {\bibinfo {volume} {91}},\ \bibinfo {pages} {125138} (\bibinfo {year} {2015})}\BibitemShut {NoStop}%
\bibitem [{\citenamefont {Morf}(1998)}]{Morf1998}%
  \BibitemOpen
  \bibfield  {author} {\bibinfo {author} {\bibfnamefont {R.~H.}\ \bibnamefont {Morf}},\ }\bibfield  {title} {\bibinfo {title} {Transition from quantum hall to compressible states in the second landau level: New light on the $\ensuremath{\nu}\phantom{\rule{0ex}{0ex}}=\phantom{\rule{0ex}{0ex}}5/2$ enigma},\ }\href {https://doi.org/10.1103/PhysRevLett.80.1505} {\bibfield  {journal} {\bibinfo  {journal} {Phys. Rev. Lett.}\ }\textbf {\bibinfo {volume} {80}},\ \bibinfo {pages} {1505--1508} (\bibinfo {year} {1998})}\BibitemShut {NoStop}%
\bibitem [{\citenamefont {Rezayi}\ and\ \citenamefont {Haldane}(2000)}]{Rezayi2000}%
  \BibitemOpen
  \bibfield  {author} {\bibinfo {author} {\bibfnamefont {E.~H.}\ \bibnamefont {Rezayi}}\ and\ \bibinfo {author} {\bibfnamefont {F.~D.~M.}\ \bibnamefont {Haldane}},\ }\bibfield  {title} {\bibinfo {title} {Incompressible paired hall state, stripe order, and the composite fermion liquid phase in half-filled landau levels},\ }\href {https://doi.org/10.1103/PhysRevLett.84.4685} {\bibfield  {journal} {\bibinfo  {journal} {Phys. Rev. Lett.}\ }\textbf {\bibinfo {volume} {84}},\ \bibinfo {pages} {4685--4688} (\bibinfo {year} {2000})}\BibitemShut {NoStop}%
\bibitem [{\citenamefont {Kang}\ \emph {et~al.}(2024)\citenamefont {Kang}, \citenamefont {Shen}, \citenamefont {Qiu}, \citenamefont {Zeng}, \citenamefont {Xia}, \citenamefont {Watanabe}, \citenamefont {Taniguchi}, \citenamefont {Shan},\ and\ \citenamefont {Mak}}]{non_Abelian_experiment}%
  \BibitemOpen
  \bibfield  {author} {\bibinfo {author} {\bibfnamefont {K.}~\bibnamefont {Kang}}, \bibinfo {author} {\bibfnamefont {B.}~\bibnamefont {Shen}}, \bibinfo {author} {\bibfnamefont {Y.}~\bibnamefont {Qiu}}, \bibinfo {author} {\bibfnamefont {Y.}~\bibnamefont {Zeng}}, \bibinfo {author} {\bibfnamefont {Z.}~\bibnamefont {Xia}}, \bibinfo {author} {\bibfnamefont {K.}~\bibnamefont {Watanabe}}, \bibinfo {author} {\bibfnamefont {T.}~\bibnamefont {Taniguchi}}, \bibinfo {author} {\bibfnamefont {J.}~\bibnamefont {Shan}},\ and\ \bibinfo {author} {\bibfnamefont {K.~F.}\ \bibnamefont {Mak}},\ }\bibfield  {title} {\bibinfo {title} {Evidence of the fractional quantum spin hall effect in moiré mote2},\ }\href {https://doi.org/10.1038/s41586-024-07214-5} {\bibfield  {journal} {\bibinfo  {journal} {Nature}\ }\textbf {\bibinfo {volume} {628}},\ \bibinfo {pages} {522--526} (\bibinfo {year} {2024})}\BibitemShut {NoStop}%
\bibitem [{\citenamefont {Reddy}\ \emph {et~al.}(2024)\citenamefont {Reddy}, \citenamefont {Paul}, \citenamefont {Abouelkomsan},\ and\ \citenamefont {Fu}}]{mr_Aidan}%
  \BibitemOpen
  \bibfield  {author} {\bibinfo {author} {\bibfnamefont {A.~P.}\ \bibnamefont {Reddy}}, \bibinfo {author} {\bibfnamefont {N.}~\bibnamefont {Paul}}, \bibinfo {author} {\bibfnamefont {A.}~\bibnamefont {Abouelkomsan}},\ and\ \bibinfo {author} {\bibfnamefont {L.}~\bibnamefont {Fu}},\ }\bibfield  {title} {\bibinfo {title} {Non-abelian fractionalization in topological minibands},\ }\href {https://doi.org/10.1103/PhysRevLett.133.166503} {\bibfield  {journal} {\bibinfo  {journal} {Phys. Rev. Lett.}\ }\textbf {\bibinfo {volume} {133}},\ \bibinfo {pages} {166503} (\bibinfo {year} {2024})}\BibitemShut {NoStop}%
\bibitem [{\citenamefont {Fujimoto}\ \emph {et~al.}(2025)\citenamefont {Fujimoto}, \citenamefont {Parker}, \citenamefont {Dong}, \citenamefont {Khalaf}, \citenamefont {Vishwanath},\ and\ \citenamefont {Ledwith}}]{2024arXiv240300856F}%
  \BibitemOpen
  \bibfield  {author} {\bibinfo {author} {\bibfnamefont {M.}~\bibnamefont {Fujimoto}}, \bibinfo {author} {\bibfnamefont {D.~E.}\ \bibnamefont {Parker}}, \bibinfo {author} {\bibfnamefont {J.}~\bibnamefont {Dong}}, \bibinfo {author} {\bibfnamefont {E.}~\bibnamefont {Khalaf}}, \bibinfo {author} {\bibfnamefont {A.}~\bibnamefont {Vishwanath}},\ and\ \bibinfo {author} {\bibfnamefont {P.}~\bibnamefont {Ledwith}},\ }\bibfield  {title} {\bibinfo {title} {Higher vortexability: Zero-field realization of higher landau levels},\ }\href {https://doi.org/10.1103/PhysRevLett.134.106502} {\bibfield  {journal} {\bibinfo  {journal} {Phys. Rev. Lett.}\ }\textbf {\bibinfo {volume} {134}},\ \bibinfo {pages} {106502} (\bibinfo {year} {2025})}\BibitemShut {NoStop}%
\bibitem [{\citenamefont {Wang}\ \emph {et~al.}(2025)\citenamefont {Wang}, \citenamefont {Zhang}, \citenamefont {Liu}, \citenamefont {Wang}, \citenamefont {Cao},\ and\ \citenamefont {Xiao}}]{mr_Wang}%
  \BibitemOpen
  \bibfield  {author} {\bibinfo {author} {\bibfnamefont {C.}~\bibnamefont {Wang}}, \bibinfo {author} {\bibfnamefont {X.-W.}\ \bibnamefont {Zhang}}, \bibinfo {author} {\bibfnamefont {X.}~\bibnamefont {Liu}}, \bibinfo {author} {\bibfnamefont {J.}~\bibnamefont {Wang}}, \bibinfo {author} {\bibfnamefont {T.}~\bibnamefont {Cao}},\ and\ \bibinfo {author} {\bibfnamefont {D.}~\bibnamefont {Xiao}},\ }\bibfield  {title} {\bibinfo {title} {Higher landau-level analogs and signatures of non-abelian states in twisted bilayer ${\mathrm{mote}}_{2}$},\ }\href {https://doi.org/10.1103/PhysRevLett.134.076503} {\bibfield  {journal} {\bibinfo  {journal} {Phys. Rev. Lett.}\ }\textbf {\bibinfo {volume} {134}},\ \bibinfo {pages} {076503} (\bibinfo {year} {2025})}\BibitemShut {NoStop}%
\bibitem [{\citenamefont {Ahn}\ \emph {et~al.}(2024)\citenamefont {Ahn}, \citenamefont {Lee}, \citenamefont {Yananose}, \citenamefont {Kim},\ and\ \citenamefont {Cho}}]{mr_Ahn}%
  \BibitemOpen
  \bibfield  {author} {\bibinfo {author} {\bibfnamefont {C.-E.}\ \bibnamefont {Ahn}}, \bibinfo {author} {\bibfnamefont {W.}~\bibnamefont {Lee}}, \bibinfo {author} {\bibfnamefont {K.}~\bibnamefont {Yananose}}, \bibinfo {author} {\bibfnamefont {Y.}~\bibnamefont {Kim}},\ and\ \bibinfo {author} {\bibfnamefont {G.~Y.}\ \bibnamefont {Cho}},\ }\bibfield  {title} {\bibinfo {title} {Non-abelian fractional quantum anomalous hall states and first landau level physics of the second moir\'e band of twisted bilayer ${\mathrm{mote}}_{2}$},\ }\href {https://doi.org/10.1103/PhysRevB.110.L161109} {\bibfield  {journal} {\bibinfo  {journal} {Phys. Rev. B}\ }\textbf {\bibinfo {volume} {110}},\ \bibinfo {pages} {L161109} (\bibinfo {year} {2024})}\BibitemShut {NoStop}%
\bibitem [{\citenamefont {Xu}\ \emph {et~al.}(2025)\citenamefont {Xu}, \citenamefont {Mao}, \citenamefont {Zeng},\ and\ \citenamefont {Zhang}}]{mr_Xu}%
  \BibitemOpen
  \bibfield  {author} {\bibinfo {author} {\bibfnamefont {C.}~\bibnamefont {Xu}}, \bibinfo {author} {\bibfnamefont {N.}~\bibnamefont {Mao}}, \bibinfo {author} {\bibfnamefont {T.}~\bibnamefont {Zeng}},\ and\ \bibinfo {author} {\bibfnamefont {Y.}~\bibnamefont {Zhang}},\ }\bibfield  {title} {\bibinfo {title} {Multiple chern bands in twisted ${\mathrm{mote}}_{2}$ and possible non-abelian states},\ }\href {https://doi.org/10.1103/PhysRevLett.134.066601} {\bibfield  {journal} {\bibinfo  {journal} {Phys. Rev. Lett.}\ }\textbf {\bibinfo {volume} {134}},\ \bibinfo {pages} {066601} (\bibinfo {year} {2025})}\BibitemShut {NoStop}%
\bibitem [{\citenamefont {Zhang}\ and\ \citenamefont {Song}(2024)}]{mr_zhang}%
  \BibitemOpen
  \bibfield  {author} {\bibinfo {author} {\bibfnamefont {L.}~\bibnamefont {Zhang}}\ and\ \bibinfo {author} {\bibfnamefont {X.-Y.}\ \bibnamefont {Song}},\ }\bibfield  {title} {\bibinfo {title} {Moore-read state in half-filled moir\'e chern band from three-body pseudopotential},\ }\href {https://doi.org/10.1103/PhysRevB.109.245128} {\bibfield  {journal} {\bibinfo  {journal} {Phys. Rev. B}\ }\textbf {\bibinfo {volume} {109}},\ \bibinfo {pages} {245128} (\bibinfo {year} {2024})}\BibitemShut {NoStop}%
\bibitem [{\citenamefont {Chen}\ \emph {et~al.}(2025)\citenamefont {Chen}, \citenamefont {Luo}, \citenamefont {Zhu},\ and\ \citenamefont {Sheng}}]{mr_Chen}%
  \BibitemOpen
  \bibfield  {author} {\bibinfo {author} {\bibfnamefont {F.}~\bibnamefont {Chen}}, \bibinfo {author} {\bibfnamefont {W.-W.}\ \bibnamefont {Luo}}, \bibinfo {author} {\bibfnamefont {W.}~\bibnamefont {Zhu}},\ and\ \bibinfo {author} {\bibfnamefont {D.~N.}\ \bibnamefont {Sheng}},\ }\bibfield  {title} {\bibinfo {title} {Robust non-{Abelian} even-denominator fractional {Chern} insulator in twisted bilayer {MoTe2}},\ }\href {https://doi.org/10.1038/s41467-025-57326-3} {\bibfield  {journal} {\bibinfo  {journal} {Nature Communications}\ }\textbf {\bibinfo {volume} {16}},\ \bibinfo {pages} {2115} (\bibinfo {year} {2025})}\BibitemShut {NoStop}%
\bibitem [{\citenamefont {Zhang}(2024)}]{yahui_zhang_non_abelian}%
  \BibitemOpen
  \bibfield  {author} {\bibinfo {author} {\bibfnamefont {Y.-H.}\ \bibnamefont {Zhang}},\ }\bibfield  {title} {\bibinfo {title} {Non-abelian and abelian descendants of a vortex spin liquid: Fractional quantum spin hall effect in twisted ${\mathrm{mote}}_{2}$},\ }\href {https://doi.org/10.1103/PhysRevB.110.155102} {\bibfield  {journal} {\bibinfo  {journal} {Phys. Rev. B}\ }\textbf {\bibinfo {volume} {110}},\ \bibinfo {pages} {155102} (\bibinfo {year} {2024})}\BibitemShut {NoStop}%
\bibitem [{\citenamefont {Sodemann~Villadiego}(2024)}]{Inti_sodemann_fractional_quantum_spin_Hall}%
  \BibitemOpen
  \bibfield  {author} {\bibinfo {author} {\bibfnamefont {I.}~\bibnamefont {Sodemann~Villadiego}},\ }\bibfield  {title} {\bibinfo {title} {Halperin states of particles and holes in ideal time reversal invariant pairs of chern bands and the fractional quantum spin hall effect in moir\'e ${\mathrm{mote}}_{2}$},\ }\href {https://doi.org/10.1103/PhysRevB.110.045114} {\bibfield  {journal} {\bibinfo  {journal} {Phys. Rev. B}\ }\textbf {\bibinfo {volume} {110}},\ \bibinfo {pages} {045114} (\bibinfo {year} {2024})}\BibitemShut {NoStop}%
\bibitem [{\citenamefont {Wilhelm}\ \emph {et~al.}(2021)\citenamefont {Wilhelm}, \citenamefont {Lang},\ and\ \citenamefont {L\"auchli}}]{Wilhelm2021}%
  \BibitemOpen
  \bibfield  {author} {\bibinfo {author} {\bibfnamefont {P.}~\bibnamefont {Wilhelm}}, \bibinfo {author} {\bibfnamefont {T.~C.}\ \bibnamefont {Lang}},\ and\ \bibinfo {author} {\bibfnamefont {A.~M.}\ \bibnamefont {L\"auchli}},\ }\bibfield  {title} {\bibinfo {title} {Interplay of fractional chern insulator and charge density wave phases in twisted bilayer graphene},\ }\href {https://doi.org/10.1103/PhysRevB.103.125406} {\bibfield  {journal} {\bibinfo  {journal} {Phys. Rev. B}\ }\textbf {\bibinfo {volume} {103}},\ \bibinfo {pages} {125406} (\bibinfo {year} {2021})}\BibitemShut {NoStop}%
\bibitem [{\citenamefont {Wilhelm}\ \emph {et~al.}(2023)\citenamefont {Wilhelm}, \citenamefont {Lang}, \citenamefont {Scheurer},\ and\ \citenamefont {Läuchli}}]{Wilhelm_2023}%
  \BibitemOpen
  \bibfield  {author} {\bibinfo {author} {\bibfnamefont {P.}~\bibnamefont {Wilhelm}}, \bibinfo {author} {\bibfnamefont {T.}~\bibnamefont {Lang}}, \bibinfo {author} {\bibfnamefont {M.}~\bibnamefont {Scheurer}},\ and\ \bibinfo {author} {\bibfnamefont {A.}~\bibnamefont {Läuchli}},\ }\bibfield  {title} {\bibinfo {title} {Non-coplanar magnetism, topological density wave order and emergent symmetry at half-integer filling of moiré chern bands},\ }\href {http://dx.doi.org/10.21468/SciPostPhys.14.3.040} {\bibfield  {journal} {\bibinfo  {journal} {SciPost Physics}\ }\textbf {\bibinfo {volume} {14}} (\bibinfo {year} {2023})}\BibitemShut {NoStop}%
\bibitem [{\citenamefont {Goldman}\ \emph {et~al.}(2023)\citenamefont {Goldman}, \citenamefont {Reddy}, \citenamefont {Paul},\ and\ \citenamefont {Fu}}]{PhysRevLett.131.136501}%
  \BibitemOpen
  \bibfield  {author} {\bibinfo {author} {\bibfnamefont {H.}~\bibnamefont {Goldman}}, \bibinfo {author} {\bibfnamefont {A.~P.}\ \bibnamefont {Reddy}}, \bibinfo {author} {\bibfnamefont {N.}~\bibnamefont {Paul}},\ and\ \bibinfo {author} {\bibfnamefont {L.}~\bibnamefont {Fu}},\ }\bibfield  {title} {\bibinfo {title} {Zero-field composite fermi liquid in twisted semiconductor bilayers},\ }\href {https://doi.org/10.1103/PhysRevLett.131.136501} {\bibfield  {journal} {\bibinfo  {journal} {Phys. Rev. Lett.}\ }\textbf {\bibinfo {volume} {131}},\ \bibinfo {pages} {136501} (\bibinfo {year} {2023})}\BibitemShut {NoStop}%
\bibitem [{\citenamefont {Dong}\ \emph {et~al.}(2023)\citenamefont {Dong}, \citenamefont {Wang}, \citenamefont {Ledwith}, \citenamefont {Vishwanath},\ and\ \citenamefont {Parker}}]{PhysRevLett.131.136502}%
  \BibitemOpen
  \bibfield  {author} {\bibinfo {author} {\bibfnamefont {J.}~\bibnamefont {Dong}}, \bibinfo {author} {\bibfnamefont {J.}~\bibnamefont {Wang}}, \bibinfo {author} {\bibfnamefont {P.~J.}\ \bibnamefont {Ledwith}}, \bibinfo {author} {\bibfnamefont {A.}~\bibnamefont {Vishwanath}},\ and\ \bibinfo {author} {\bibfnamefont {D.~E.}\ \bibnamefont {Parker}},\ }\bibfield  {title} {\bibinfo {title} {Composite fermi liquid at zero magnetic field in twisted ${\mathrm{mote}}_{2}$},\ }\href {https://doi.org/10.1103/PhysRevLett.131.136502} {\bibfield  {journal} {\bibinfo  {journal} {Phys. Rev. Lett.}\ }\textbf {\bibinfo {volume} {131}},\ \bibinfo {pages} {136502} (\bibinfo {year} {2023})}\BibitemShut {NoStop}%
\bibitem [{\citenamefont {Tarnopolsky}\ \emph {et~al.}(2019)\citenamefont {Tarnopolsky}, \citenamefont {Kruchkov},\ and\ \citenamefont {Vishwanath}}]{PhysRevLett.122.106405}%
  \BibitemOpen
  \bibfield  {author} {\bibinfo {author} {\bibfnamefont {G.}~\bibnamefont {Tarnopolsky}}, \bibinfo {author} {\bibfnamefont {A.~J.}\ \bibnamefont {Kruchkov}},\ and\ \bibinfo {author} {\bibfnamefont {A.}~\bibnamefont {Vishwanath}},\ }\bibfield  {title} {\bibinfo {title} {Origin of magic angles in twisted bilayer graphene},\ }\href {https://doi.org/10.1103/PhysRevLett.122.106405} {\bibfield  {journal} {\bibinfo  {journal} {Phys. Rev. Lett.}\ }\textbf {\bibinfo {volume} {122}},\ \bibinfo {pages} {106405} (\bibinfo {year} {2019})}\BibitemShut {NoStop}%
\bibitem [{Sup()}]{SupMat}%
  \BibitemOpen
  \href@noop {} {\bibinfo  {journal} {In this supplemental material, we provide more numerical data of the single-particle band dispersion, valley polarization simulations, Berry curvature and quantum metric of the flat band, single-hole energy of moiré minibands and Landau levels, chiral graviton spectral function, spectral flow of the Moore-Read states, the composite Fermi liquid at weak coupling strength, and results on charge density waves at larger system sizes. We also show the energy gap scaling of Moore-Read states and the derivation of quasihole counting.}\ }\BibitemShut {NoStop}%
\bibitem [{\citenamefont {Xu}\ \emph {et~al.}(2024)\citenamefont {Xu}, \citenamefont {Li}, \citenamefont {Xu}, \citenamefont {Bi},\ and\ \citenamefont {Zhang}}]{xu2024maximally}%
  \BibitemOpen
\bibfield  {journal} {  }\bibfield  {author} {\bibinfo {author} {\bibfnamefont {C.}~\bibnamefont {Xu}}, \bibinfo {author} {\bibfnamefont {J.}~\bibnamefont {Li}}, \bibinfo {author} {\bibfnamefont {Y.}~\bibnamefont {Xu}}, \bibinfo {author} {\bibfnamefont {Z.}~\bibnamefont {Bi}},\ and\ \bibinfo {author} {\bibfnamefont {Y.}~\bibnamefont {Zhang}},\ }\bibfield  {title} {\bibinfo {title} {Maximally localized wannier functions, interaction models, and fractional quantum anomalous hall effect in twisted bilayer mote 2},\ }\bibfield  {journal} {\bibinfo  {journal} {Proceedings of the National Academy of Sciences}\ }\textbf {\bibinfo {volume} {121}},\ \href {https://doi.org/10.1073/pnas.2316749121} {10.1073/pnas.2316749121} (\bibinfo {year} {2024})\BibitemShut {NoStop}%
\bibitem [{\citenamefont {Yu}\ \emph {et~al.}(2024)\citenamefont {Yu}, \citenamefont {Herzog-Arbeitman}, \citenamefont {Wang}, \citenamefont {Vafek}, \citenamefont {Bernevig},\ and\ \citenamefont {Regnault}}]{yu2024fractional}%
  \BibitemOpen
  \bibfield  {author} {\bibinfo {author} {\bibfnamefont {J.}~\bibnamefont {Yu}}, \bibinfo {author} {\bibfnamefont {J.}~\bibnamefont {Herzog-Arbeitman}}, \bibinfo {author} {\bibfnamefont {M.}~\bibnamefont {Wang}}, \bibinfo {author} {\bibfnamefont {O.}~\bibnamefont {Vafek}}, \bibinfo {author} {\bibfnamefont {B.~A.}\ \bibnamefont {Bernevig}},\ and\ \bibinfo {author} {\bibfnamefont {N.}~\bibnamefont {Regnault}},\ }\bibfield  {title} {\bibinfo {title} {Fractional {{Chern}} insulators versus nonmagnetic states in twisted bilayer ${\mathrm{mote}}_{2}$},\ }\href {https://doi.org/10.1103/PhysRevB.109.045147} {\bibfield  {journal} {\bibinfo  {journal} {Phys. Rev. B}\ }\textbf {\bibinfo {volume} {109}},\ \bibinfo {pages} {045147} (\bibinfo {year} {2024})}\BibitemShut {NoStop}%
\bibitem [{\citenamefont {Abouelkomsan}\ \emph {et~al.}(2024)\citenamefont {Abouelkomsan}, \citenamefont {Reddy}, \citenamefont {Fu},\ and\ \citenamefont {Bergholtz}}]{abouelkomsan2024band}%
  \BibitemOpen
  \bibfield  {author} {\bibinfo {author} {\bibfnamefont {A.}~\bibnamefont {Abouelkomsan}}, \bibinfo {author} {\bibfnamefont {A.~P.}\ \bibnamefont {Reddy}}, \bibinfo {author} {\bibfnamefont {L.}~\bibnamefont {Fu}},\ and\ \bibinfo {author} {\bibfnamefont {E.~J.}\ \bibnamefont {Bergholtz}},\ }\bibfield  {title} {\bibinfo {title} {Band mixing in the quantum anomalous hall regime of twisted semiconductor bilayers},\ }\href {https://doi.org/10.1103/PhysRevB.109.L121107} {\bibfield  {journal} {\bibinfo  {journal} {Phys. Rev. B}\ }\textbf {\bibinfo {volume} {109}},\ \bibinfo {pages} {L121107} (\bibinfo {year} {2024})}\BibitemShut {NoStop}%
\bibitem [{\citenamefont {Bergholtz}\ \emph {et~al.}(2006)\citenamefont {Bergholtz}, \citenamefont {Kailasvuori}, \citenamefont {Wikberg}, \citenamefont {Hansson},\ and\ \citenamefont {Karlhede}}]{Bergholtz2006}%
  \BibitemOpen
  \bibfield  {author} {\bibinfo {author} {\bibfnamefont {E.~J.}\ \bibnamefont {Bergholtz}}, \bibinfo {author} {\bibfnamefont {J.}~\bibnamefont {Kailasvuori}}, \bibinfo {author} {\bibfnamefont {E.}~\bibnamefont {Wikberg}}, \bibinfo {author} {\bibfnamefont {T.~H.}\ \bibnamefont {Hansson}},\ and\ \bibinfo {author} {\bibfnamefont {A.}~\bibnamefont {Karlhede}},\ }\bibfield  {title} {\bibinfo {title} {Pfaffian quantum hall state made simple: Multiple vacua and domain walls on a thin torus},\ }\href {https://doi.org/10.1103/PhysRevB.74.081308} {\bibfield  {journal} {\bibinfo  {journal} {Phys. Rev. B}\ }\textbf {\bibinfo {volume} {74}},\ \bibinfo {pages} {081308} (\bibinfo {year} {2006})}\BibitemShut {NoStop}%
\bibitem [{\citenamefont {Seidel}\ and\ \citenamefont {Lee}(2006)}]{Seidel2006}%
  \BibitemOpen
  \bibfield  {author} {\bibinfo {author} {\bibfnamefont {A.}~\bibnamefont {Seidel}}\ and\ \bibinfo {author} {\bibfnamefont {D.-H.}\ \bibnamefont {Lee}},\ }\bibfield  {title} {\bibinfo {title} {Abelian and non-abelian hall liquids and charge-density wave: Quantum number fractionalization in one and two dimensions},\ }\href {https://doi.org/10.1103/PhysRevLett.97.056804} {\bibfield  {journal} {\bibinfo  {journal} {Phys. Rev. Lett.}\ }\textbf {\bibinfo {volume} {97}},\ \bibinfo {pages} {056804} (\bibinfo {year} {2006})}\BibitemShut {NoStop}%
\bibitem [{\citenamefont {Ardonne}\ \emph {et~al.}(2008)\citenamefont {Ardonne}, \citenamefont {Bergholtz}, \citenamefont {Kailasvuori},\ and\ \citenamefont {Wikberg}}]{Ardonne_2008}%
  \BibitemOpen
  \bibfield  {author} {\bibinfo {author} {\bibfnamefont {E.}~\bibnamefont {Ardonne}}, \bibinfo {author} {\bibfnamefont {E.~J.}\ \bibnamefont {Bergholtz}}, \bibinfo {author} {\bibfnamefont {J.}~\bibnamefont {Kailasvuori}},\ and\ \bibinfo {author} {\bibfnamefont {E.}~\bibnamefont {Wikberg}},\ }\bibfield  {title} {\bibinfo {title} {Degeneracy of non-abelian quantum hall states on the torus: domain walls and conformal field theory},\ }\href {https://doi.org/10.1088/1742-5468/2008/04/P04016} {\bibfield  {journal} {\bibinfo  {journal} {Journal of Statistical Mechanics: Theory and Experiment}\ }\textbf {\bibinfo {volume} {2008}},\ \bibinfo {pages} {P04016} (\bibinfo {year} {2008})}\BibitemShut {NoStop}%
\bibitem [{\citenamefont {Read}(2006)}]{PhysRevB.73.245334}%
  \BibitemOpen
  \bibfield  {author} {\bibinfo {author} {\bibfnamefont {N.}~\bibnamefont {Read}},\ }\bibfield  {title} {\bibinfo {title} {Wavefunctions and counting formulas for quasiholes of clustered quantum hall states on a sphere},\ }\href {https://doi.org/10.1103/PhysRevB.73.245334} {\bibfield  {journal} {\bibinfo  {journal} {Phys. Rev. B}\ }\textbf {\bibinfo {volume} {73}},\ \bibinfo {pages} {245334} (\bibinfo {year} {2006})}\BibitemShut {NoStop}%
\bibitem [{\citenamefont {Li}\ and\ \citenamefont {Haldane}(2008)}]{li_entanglement_2008}%
  \BibitemOpen
  \bibfield  {author} {\bibinfo {author} {\bibfnamefont {H.}~\bibnamefont {Li}}\ and\ \bibinfo {author} {\bibfnamefont {F.~D.~M.}\ \bibnamefont {Haldane}},\ }\bibfield  {title} {\bibinfo {title} {Entanglement {Spectrum} as a {Generalization} of {Entanglement} {Entropy}: {Identification} of {Topological} {Order} in {Non}-{Abelian} {Fractional} {Quantum} {Hall} {Effect} {States}},\ }\href {https://doi.org/10.1103/PhysRevLett.101.010504} {\bibfield  {journal} {\bibinfo  {journal} {Physical Review Letters}\ }\textbf {\bibinfo {volume} {101}},\ \bibinfo {pages} {010504} (\bibinfo {year} {2008})}\BibitemShut {NoStop}%
\bibitem [{\citenamefont {Sterdyniak}\ \emph {et~al.}(2011)\citenamefont {Sterdyniak}, \citenamefont {Regnault},\ and\ \citenamefont {Bernevig}}]{sterdyniak_extracting_2011}%
  \BibitemOpen
  \bibfield  {author} {\bibinfo {author} {\bibfnamefont {A.}~\bibnamefont {Sterdyniak}}, \bibinfo {author} {\bibfnamefont {N.}~\bibnamefont {Regnault}},\ and\ \bibinfo {author} {\bibfnamefont {B.~A.}\ \bibnamefont {Bernevig}},\ }\bibfield  {title} {\bibinfo {title} {Extracting {Excitations} from {Model} {State} {Entanglement}},\ }\href {https://doi.org/10.1103/PhysRevLett.106.100405} {\bibfield  {journal} {\bibinfo  {journal} {Physical Review Letters}\ }\textbf {\bibinfo {volume} {106}},\ \bibinfo {pages} {100405} (\bibinfo {year} {2011})}\BibitemShut {NoStop}%
\bibitem [{\citenamefont {{Bernevig}}\ and\ \citenamefont {{Regnault}}(2012)}]{2012arXiv1204.5682B}%
  \BibitemOpen
  \bibfield  {author} {\bibinfo {author} {\bibfnamefont {B.~A.}\ \bibnamefont {{Bernevig}}}\ and\ \bibinfo {author} {\bibfnamefont {N.}~\bibnamefont {{Regnault}}},\ }\bibfield  {title} {\bibinfo {title} {{Thin-Torus Limit of Fractional Topological Insulators}},\ }\href {https://doi.org/10.48550/arXiv.1204.5682} {\bibfield  {journal} {\bibinfo  {journal} {arXiv e-prints}\ ,\ \bibinfo {eid} {arXiv:1204.5682}} (\bibinfo {year} {2012})},\ \Eprint {https://arxiv.org/abs/1204.5682} {arXiv:1204.5682 [cond-mat.str-el]} \BibitemShut {NoStop}%
\bibitem [{\citenamefont {Yoshioka}\ \emph {et~al.}(1983)\citenamefont {Yoshioka}, \citenamefont {Halperin},\ and\ \citenamefont {Lee}}]{Yoshioka1983}%
  \BibitemOpen
  \bibfield  {author} {\bibinfo {author} {\bibfnamefont {D.}~\bibnamefont {Yoshioka}}, \bibinfo {author} {\bibfnamefont {B.~I.}\ \bibnamefont {Halperin}},\ and\ \bibinfo {author} {\bibfnamefont {P.~A.}\ \bibnamefont {Lee}},\ }\bibfield  {title} {\bibinfo {title} {Ground state of two-dimensional electrons in strong magnetic fields and $\frac{1}{3}$ quantized hall effect},\ }\href {https://doi.org/10.1103/PhysRevLett.50.1219} {\bibfield  {journal} {\bibinfo  {journal} {Phys. Rev. Lett.}\ }\textbf {\bibinfo {volume} {50}},\ \bibinfo {pages} {1219--1222} (\bibinfo {year} {1983})}\BibitemShut {NoStop}%
\bibitem [{\citenamefont {Abouelkomsan}\ \emph {et~al.}(2023)\citenamefont {Abouelkomsan}, \citenamefont {Yang},\ and\ \citenamefont {Bergholtz}}]{PhysRevResearch.5.L012015}%
  \BibitemOpen
  \bibfield  {author} {\bibinfo {author} {\bibfnamefont {A.}~\bibnamefont {Abouelkomsan}}, \bibinfo {author} {\bibfnamefont {K.}~\bibnamefont {Yang}},\ and\ \bibinfo {author} {\bibfnamefont {E.~J.}\ \bibnamefont {Bergholtz}},\ }\bibfield  {title} {\bibinfo {title} {Quantum metric induced phases in moir\'e materials},\ }\href {https://doi.org/10.1103/PhysRevResearch.5.L012015} {\bibfield  {journal} {\bibinfo  {journal} {Phys. Rev. Res.}\ }\textbf {\bibinfo {volume} {5}},\ \bibinfo {pages} {L012015} (\bibinfo {year} {2023})}\BibitemShut {NoStop}%
\bibitem [{\citenamefont {{Ji}}\ and\ \citenamefont {{Yang}}(2024)}]{2024arXiv240908324J}%
  \BibitemOpen
  \bibfield  {author} {\bibinfo {author} {\bibfnamefont {G.}~\bibnamefont {{Ji}}}\ and\ \bibinfo {author} {\bibfnamefont {B.}~\bibnamefont {{Yang}}},\ }\bibfield  {title} {\bibinfo {title} {{Quantum metric induced hole dispersion and emergent particle-hole symmetry in topological flat bands}},\ }\href {https://doi.org/10.48550/arXiv.2409.08324} {\bibfield  {journal} {\bibinfo  {journal} {arXiv e-prints}\ ,\ \bibinfo {eid} {arXiv:2409.08324}} (\bibinfo {year} {2024})},\ \Eprint {https://arxiv.org/abs/2409.08324} {arXiv:2409.08324 [cond-mat.str-el]} \BibitemShut {NoStop}%
\bibitem [{\citenamefont {Liu}\ \emph {et~al.}(2025)\citenamefont {Liu}, \citenamefont {Mera}, \citenamefont {Fujimoto}, \citenamefont {Ozawa},\ and\ \citenamefont {Wang}}]{liu2024theorygeneralizedlandaulevels}%
  \BibitemOpen
  \bibfield  {author} {\bibinfo {author} {\bibfnamefont {Z.}~\bibnamefont {Liu}}, \bibinfo {author} {\bibfnamefont {B.}~\bibnamefont {Mera}}, \bibinfo {author} {\bibfnamefont {M.}~\bibnamefont {Fujimoto}}, \bibinfo {author} {\bibfnamefont {T.}~\bibnamefont {Ozawa}},\ and\ \bibinfo {author} {\bibfnamefont {J.}~\bibnamefont {Wang}},\ }\bibfield  {title} {\bibinfo {title} {Theory of generalized landau levels and its implications for non-abelian states},\ }\href {https://doi.org/10.1103/1zg9-qbd6} {\bibfield  {journal} {\bibinfo  {journal} {Phys. Rev. X}\ }\textbf {\bibinfo {volume} {15}},\ \bibinfo {pages} {031019} (\bibinfo {year} {2025})}\BibitemShut {NoStop}%
\bibitem [{\citenamefont {Yang}(2013)}]{form_factor_ll}%
  \BibitemOpen
  \bibfield  {author} {\bibinfo {author} {\bibfnamefont {K.}~\bibnamefont {Yang}},\ }\bibfield  {title} {\bibinfo {title} {Geometry of compressible and incompressible quantum hall states: Application to anisotropic composite-fermion liquids},\ }\href {https://doi.org/10.1103/PhysRevB.88.241105} {\bibfield  {journal} {\bibinfo  {journal} {Phys. Rev. B}\ }\textbf {\bibinfo {volume} {88}},\ \bibinfo {pages} {241105} (\bibinfo {year} {2013})}\BibitemShut {NoStop}%
\bibitem [{\citenamefont {Nguyen}\ and\ \citenamefont {Son}(2021)}]{graviton0}%
  \BibitemOpen
  \bibfield  {author} {\bibinfo {author} {\bibfnamefont {D.~X.}\ \bibnamefont {Nguyen}}\ and\ \bibinfo {author} {\bibfnamefont {D.~T.}\ \bibnamefont {Son}},\ }\bibfield  {title} {\bibinfo {title} {Probing the spin structure of the fractional quantum {Hall} magnetoroton with polarized raman scattering},\ }\href {https://doi.org/10.1103/PhysRevResearch.3.023040} {\bibfield  {journal} {\bibinfo  {journal} {Phys. Rev. Res.}\ }\textbf {\bibinfo {volume} {3}},\ \bibinfo {pages} {023040} (\bibinfo {year} {2021})}\BibitemShut {NoStop}%
\bibitem [{\citenamefont {{Nguyen}}\ \emph {et~al.}(2014)\citenamefont {{Nguyen}}, \citenamefont {{Thanh Son}},\ and\ \citenamefont {{Wu}}}]{graviton00}%
  \BibitemOpen
  \bibfield  {author} {\bibinfo {author} {\bibfnamefont {D.~X.}\ \bibnamefont {{Nguyen}}}, \bibinfo {author} {\bibfnamefont {D.}~\bibnamefont {{Thanh Son}}},\ and\ \bibinfo {author} {\bibfnamefont {C.}~\bibnamefont {{Wu}}},\ }\bibfield  {title} {\bibinfo {title} {{Lowest Landau Level Stress Tensor and Structure Factor of Trial Quantum Hall Wave Functions}},\ }\href {https://doi.org/10.48550/arXiv.1411.3316} {\bibfield  {journal} {\bibinfo  {journal} {arXiv e-prints}\ ,\ \bibinfo {eid} {arXiv:1411.3316}} (\bibinfo {year} {2014})},\ \Eprint {https://arxiv.org/abs/1411.3316} {arXiv:1411.3316 [cond-mat.str-el]} \BibitemShut {NoStop}%
\bibitem [{\citenamefont {Liou}\ \emph {et~al.}(2019)\citenamefont {Liou}, \citenamefont {Haldane}, \citenamefont {Yang},\ and\ \citenamefont {Rezayi}}]{graviton000}%
  \BibitemOpen
  \bibfield  {author} {\bibinfo {author} {\bibfnamefont {S.-F.}\ \bibnamefont {Liou}}, \bibinfo {author} {\bibfnamefont {F.~D.~M.}\ \bibnamefont {Haldane}}, \bibinfo {author} {\bibfnamefont {K.}~\bibnamefont {Yang}},\ and\ \bibinfo {author} {\bibfnamefont {E.~H.}\ \bibnamefont {Rezayi}},\ }\bibfield  {title} {\bibinfo {title} {Chiral gravitons in fractional quantum {Hall} liquids},\ }\href {https://doi.org/10.1103/PhysRevLett.123.146801} {\bibfield  {journal} {\bibinfo  {journal} {Phys. Rev. Lett.}\ }\textbf {\bibinfo {volume} {123}},\ \bibinfo {pages} {146801} (\bibinfo {year} {2019})}\BibitemShut {NoStop}%
\bibitem [{\citenamefont {{Shen}}\ \emph {et~al.}(2024)\citenamefont {{Shen}}, \citenamefont {{Wang}}, \citenamefont {{Hu}}, \citenamefont {{Guo}}, \citenamefont {{Yao}}, \citenamefont {{Wang}}, \citenamefont {{Duan}},\ and\ \citenamefont {{Xu}}}]{graviton1}%
  \BibitemOpen
  \bibfield  {author} {\bibinfo {author} {\bibfnamefont {X.}~\bibnamefont {{Shen}}}, \bibinfo {author} {\bibfnamefont {C.}~\bibnamefont {{Wang}}}, \bibinfo {author} {\bibfnamefont {X.}~\bibnamefont {{Hu}}}, \bibinfo {author} {\bibfnamefont {R.}~\bibnamefont {{Guo}}}, \bibinfo {author} {\bibfnamefont {H.}~\bibnamefont {{Yao}}}, \bibinfo {author} {\bibfnamefont {C.}~\bibnamefont {{Wang}}}, \bibinfo {author} {\bibfnamefont {W.}~\bibnamefont {{Duan}}},\ and\ \bibinfo {author} {\bibfnamefont {Y.}~\bibnamefont {{Xu}}},\ }\bibfield  {title} {\bibinfo {title} {{Magnetorotons in Moir{\'e} Fractional Chern Insulators}},\ }\href {https://doi.org/10.48550/arXiv.2412.01211} {\bibfield  {journal} {\bibinfo  {journal} {arXiv e-prints}\ ,\ \bibinfo {eid} {arXiv:2412.01211}} (\bibinfo {year} {2024})},\ \Eprint {https://arxiv.org/abs/2412.01211} {arXiv:2412.01211 [cond-mat.str-el]} \BibitemShut {NoStop}%
\bibitem [{\citenamefont {{Long}}\ \emph {et~al.}(2024)\citenamefont {{Long}}, \citenamefont {{Lu}}, \citenamefont {{Wu}},\ and\ \citenamefont {{Meng}}}]{graviton2}%
  \BibitemOpen
  \bibfield  {author} {\bibinfo {author} {\bibfnamefont {M.}~\bibnamefont {{Long}}}, \bibinfo {author} {\bibfnamefont {H.}~\bibnamefont {{Lu}}}, \bibinfo {author} {\bibfnamefont {H.-Q.}\ \bibnamefont {{Wu}}},\ and\ \bibinfo {author} {\bibfnamefont {Z.~Y.}\ \bibnamefont {{Meng}}},\ }\bibfield  {title} {\bibinfo {title} {{Spectra of Magnetoroton and Chiral Graviton Modes of Fractional Chern Insulator}},\ }\href {https://doi.org/10.48550/arXiv.2501.00247} {\bibfield  {journal} {\bibinfo  {journal} {arXiv e-prints}\ ,\ \bibinfo {eid} {arXiv:2501.00247}} (\bibinfo {year} {2024})},\ \Eprint {https://arxiv.org/abs/2501.00247} {arXiv:2501.00247 [cond-mat.str-el]} \BibitemShut {NoStop}%
\bibitem [{\citenamefont {{Wang}}\ \emph {et~al.}(2025)\citenamefont {{Wang}}, \citenamefont {{Huxford}}, \citenamefont {{Nguyen}}, \citenamefont {{Ji}}, \citenamefont {{Kim}},\ and\ \citenamefont {{Yang}}}]{graviton3}%
  \BibitemOpen
  \bibfield  {author} {\bibinfo {author} {\bibfnamefont {Y.}~\bibnamefont {{Wang}}}, \bibinfo {author} {\bibfnamefont {J.}~\bibnamefont {{Huxford}}}, \bibinfo {author} {\bibfnamefont {D.~X.}\ \bibnamefont {{Nguyen}}}, \bibinfo {author} {\bibfnamefont {G.}~\bibnamefont {{Ji}}}, \bibinfo {author} {\bibfnamefont {Y.~B.}\ \bibnamefont {{Kim}}},\ and\ \bibinfo {author} {\bibfnamefont {B.}~\bibnamefont {{Yang}}},\ }\bibfield  {title} {\bibinfo {title} {{Dynamics and lifetime of geometric excitations in moir{\'e} systems}},\ }\href {https://doi.org/10.48550/arXiv.2502.02640} {\bibfield  {journal} {\bibinfo  {journal} {arXiv e-prints}\ ,\ \bibinfo {eid} {arXiv:2502.02640}} (\bibinfo {year} {2025})},\ \Eprint {https://arxiv.org/abs/2502.02640} {arXiv:2502.02640 [cond-mat.str-el]} \BibitemShut {NoStop}%
\bibitem [{\citenamefont {Haldane}\ \emph {et~al.}(2021)\citenamefont {Haldane}, \citenamefont {Rezayi},\ and\ \citenamefont {Yang}}]{moore_read_graviton}%
  \BibitemOpen
  \bibfield  {author} {\bibinfo {author} {\bibfnamefont {F.~D.~M.}\ \bibnamefont {Haldane}}, \bibinfo {author} {\bibfnamefont {E.~H.}\ \bibnamefont {Rezayi}},\ and\ \bibinfo {author} {\bibfnamefont {K.}~\bibnamefont {Yang}},\ }\bibfield  {title} {\bibinfo {title} {Graviton chirality and topological order in the half-filled landau level},\ }\href {https://doi.org/10.1103/PhysRevB.104.L121106} {\bibfield  {journal} {\bibinfo  {journal} {Phys. Rev. B}\ }\textbf {\bibinfo {volume} {104}},\ \bibinfo {pages} {L121106} (\bibinfo {year} {2021})}\BibitemShut {NoStop}%
\bibitem [{\citenamefont {Jain}(1989)}]{Jain_cfl}%
  \BibitemOpen
  \bibfield  {author} {\bibinfo {author} {\bibfnamefont {J.~K.}\ \bibnamefont {Jain}},\ }\bibfield  {title} {\bibinfo {title} {Composite-fermion approach for the fractional quantum hall effect},\ }\href {https://doi.org/10.1103/PhysRevLett.63.199} {\bibfield  {journal} {\bibinfo  {journal} {Phys. Rev. Lett.}\ }\textbf {\bibinfo {volume} {63}},\ \bibinfo {pages} {199--202} (\bibinfo {year} {1989})}\BibitemShut {NoStop}%
\bibitem [{\citenamefont {Geraedts}\ \emph {et~al.}(2016)\citenamefont {Geraedts}, \citenamefont {Zaletel}, \citenamefont {Mong}, \citenamefont {Metlitski}, \citenamefont {Vishwanath},\ and\ \citenamefont {Motrunich}}]{composite_fermion_science}%
  \BibitemOpen
  \bibfield  {author} {\bibinfo {author} {\bibfnamefont {S.~D.}\ \bibnamefont {Geraedts}}, \bibinfo {author} {\bibfnamefont {M.~P.}\ \bibnamefont {Zaletel}}, \bibinfo {author} {\bibfnamefont {R.~S.~K.}\ \bibnamefont {Mong}}, \bibinfo {author} {\bibfnamefont {M.~A.}\ \bibnamefont {Metlitski}}, \bibinfo {author} {\bibfnamefont {A.}~\bibnamefont {Vishwanath}},\ and\ \bibinfo {author} {\bibfnamefont {O.~I.}\ \bibnamefont {Motrunich}},\ }\bibfield  {title} {\bibinfo {title} {The half-filled landau level: The case for dirac composite fermions},\ }\href {https://doi.org/10.1126/science.aad4302} {\bibfield  {journal} {\bibinfo  {journal} {Science}\ }\textbf {\bibinfo {volume} {352}},\ \bibinfo {pages} {197--201} (\bibinfo {year} {2016})},\ \Eprint {https://arxiv.org/abs/https://www.science.org/doi/pdf/10.1126/science.aad4302} {https://www.science.org/doi/pdf/10.1126/science.aad4302} \BibitemShut {NoStop}%
\end{thebibliography}%

\section{end matter}
\label{sec:end_matter}
\emph{Transition between CDWs and Moore-Read states.} ---  
Having provided evidence for both the CDW and the MR states, we now revisit and provide more details on the phase transition between them by tuning $\gamma$ as was already shown in Fig. \ref{fig:transition}.   
Within the range $\gamma\in (0, 10]$, the two layers of TBGs remain coupled and the targeted flat band remains isolated.
We identify the phase transition between the CDW order and the MR states through two criteria. 
One is the many-body energy spectrum gap $\Delta_E$ between the degenerated ground states and excited states. 
Here, we choose odd number of electrons to clearly show the phase transition. (In even-electron case, the gap is less conclusive as only two states evolve to CDW orders). It then reveals a phase transition around $\gamma\approx 4.5$.  
Our second tool is the PES gap at the CDW and MR countings. 
Here, the PES gap is defined as $\Delta_{\text{PES}}=\xi_{m+1}-\xi_{m}$, where $\xi_{m}$ represents the $m$-th PES level in ascending order, with $m$ the counting for either the CDW or the MR states. We find
$\Delta_{\text{PES}}$ shows a consistent behavior as $\Delta_E$. 
Before the phase transition, the PES gap for the CDW is roughly $0$, while it is significant for the MR states (the largest $\Delta_{\text{PES}}^{\text{MR}}\approx 0.6$), which serves as a supplementary confirmation of MR states. 
Upon crossing the phase transition point, we observe a rapid decrease of $\Delta_{\text{PES}}^{\text{MR}}$ to $0$, while $\Delta_{\text{PES}}^{\text{CDW}}$ undergoes a pronounced jump, attaining a finite value.
This serves as compelling evidence for the phase transition from MR states to the CDW. 

\pagebreak
\newpage

\onecolumngrid
\pagebreak
\newpage

\section{Supplemental Materials to `Non-Abelian Fractional Chern Insulators and Competing States in Flat Moir\'e Bands'}

In this supplemental material, we provide more numerical data of the single-particle band dispersion, valley polarization simulations, Berry curvature and quantum metric of the flat band, single-hole energy of moiré minibands and Landau levels, chiral graviton spectral function, spectral flow of the Moore-Read states, the composite Fermi liquid at weak coupling strength, and results on charge density waves at larger system sizes. We also show the energy gap scaling of Moore-Read states and the derivation of quasihole counting.

\section{Band dispersion} 

\begin{figure}
\centering
\includegraphics[width=0.5\linewidth]{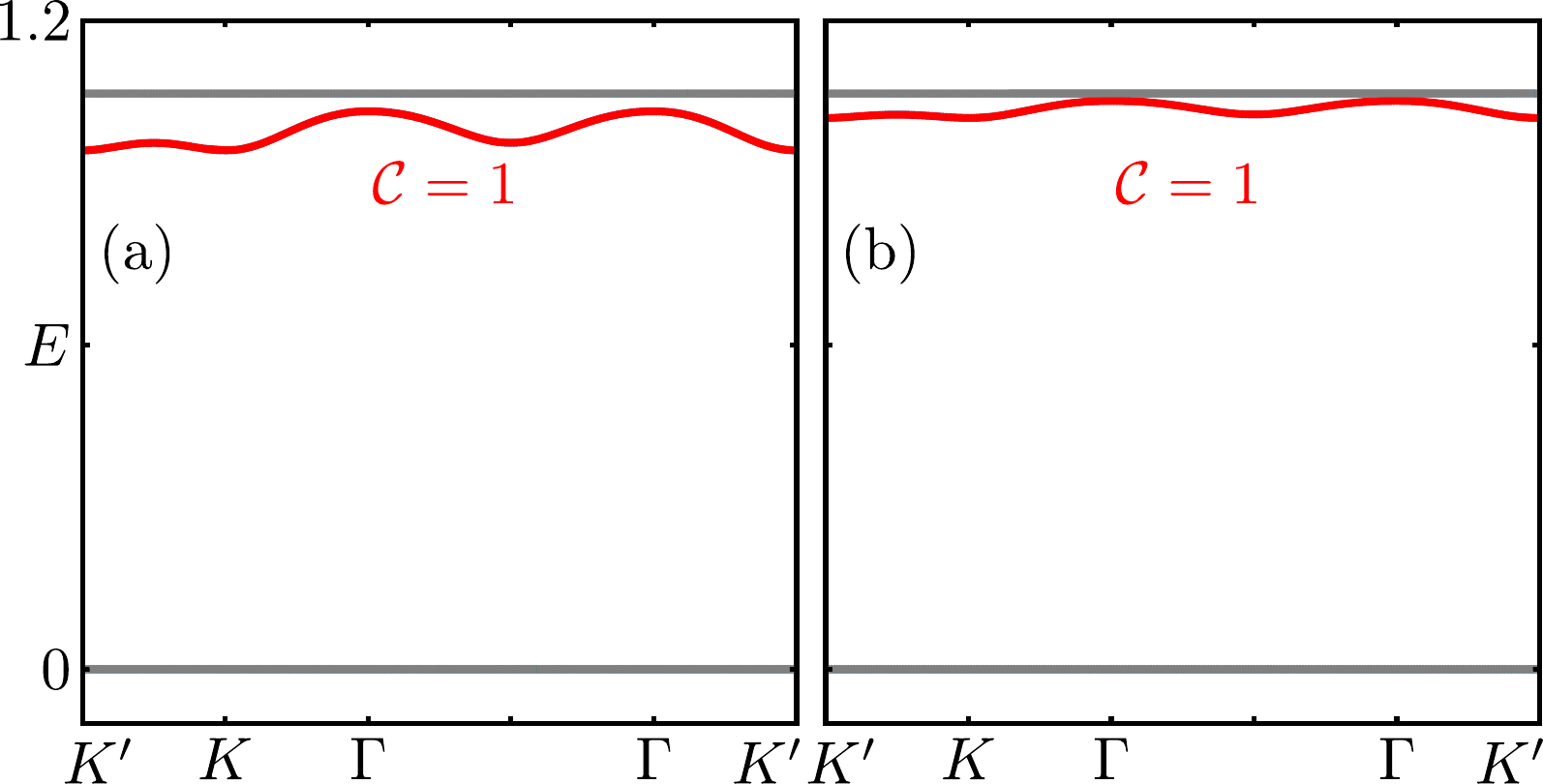}
\caption{Band dispersion of the double TBG model with $\gamma=3.5\ {\rm eV}$ (a) and $\gamma=5.5 \ {\rm eV}$ (b), respectively. Here, the red color indicates the target band and we use a cutoff $m=5$ in the low-energy valley and spin polarized Hamiltonian.
\label{fig:dispersion}
}
\end{figure}

In this section, we show the single-particle band dispersion at two values of $\gamma$ supporting the Moore-Read states and the charge density wave, respectively (see Fig.~\ref{fig:dispersion}). 
In both cases, the targeted nearly flat band remains isolated and has a Chern number $\mathcal{C}=1$.

\section{Valley polarization simulations}

\begin{figure}
\centering
\includegraphics[width=1\linewidth]{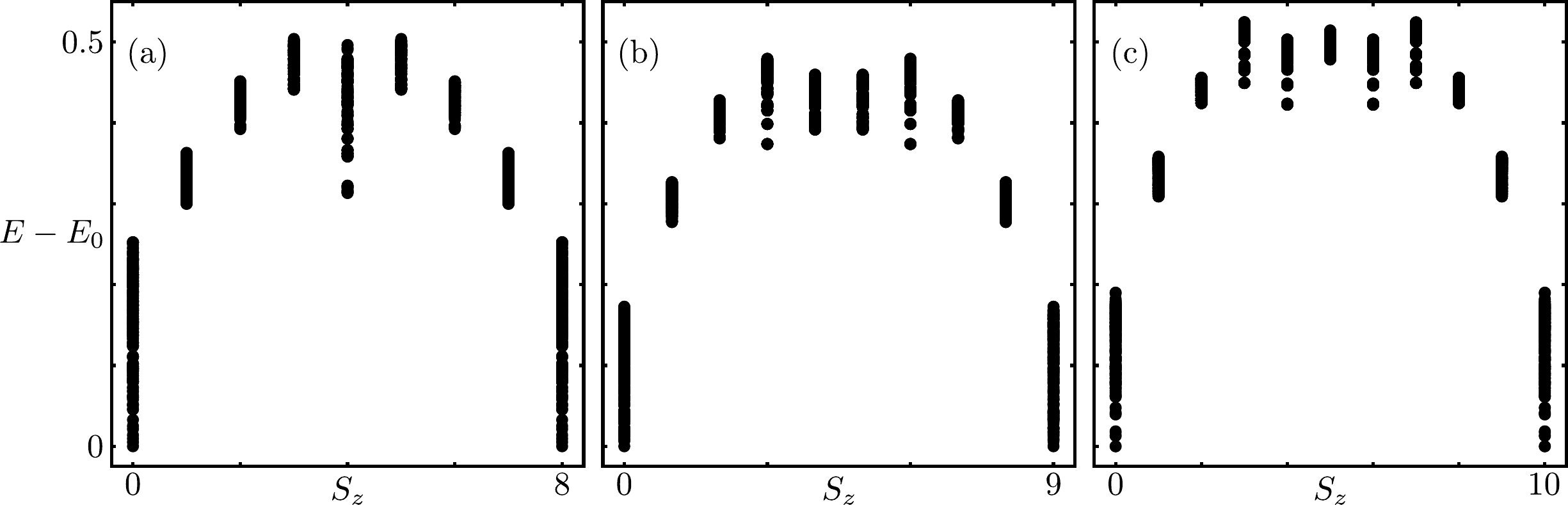}
\caption{Low-lying many-body energy spectrum at half filling, including both valleys with $N_e=8$ (a), $N_e=9$ (b), and $N_e=10$ (c), respectively. Here, $E_0$ denotes the ground state energy, and $S_z$ ($N_e-S_z$) indicates the number of electrons occupying valley $K$ ($K'$), serving as a pseudospin sector label. All calculations are performed with $\gamma=3.5$eV and on clusters with $16$, $18$, and $20$ sites per valley, respectively.
\label{fig:valley_polarization}
}
\end{figure}

In this section, we present numerical results supporting the valley polarization assumption. Here, we incorporate the single-particle information for both $K$ and $K'$ valleys. The single-particle Hamiltonian of the $K'$ valley is related to that of the $K$ valley by time-reversal symmetry. Using this, we then project the electron-electron interaction into these time-reversal-symmetry-related targeted bands.

As shown in Fig.~\ref{fig:valley_polarization}, at half filling and across different system sizes, we find that the lowest-energy states are fully polarized in either the $K$ or $K'$ valley, and are well separated from higher-energy states that exhibit mixed-valley behaviors. The presence of a sizable gap indicates that the valley-polarized states are energetically favored in the parameter regime we study, thus validating our valley polarization assumption.

We note that in more complex scenarios---such as at different fillings, with modified interactions, or in the presence of substrate effects---this assumption may no longer hold. In this sense, a systematic investigation of phases beyond the valley assumption represents an important and interesting direction, which we plan to explore in the future.  

\section{geometric tensor perspective}

\begin{figure}
\centering
\includegraphics[width=\linewidth]{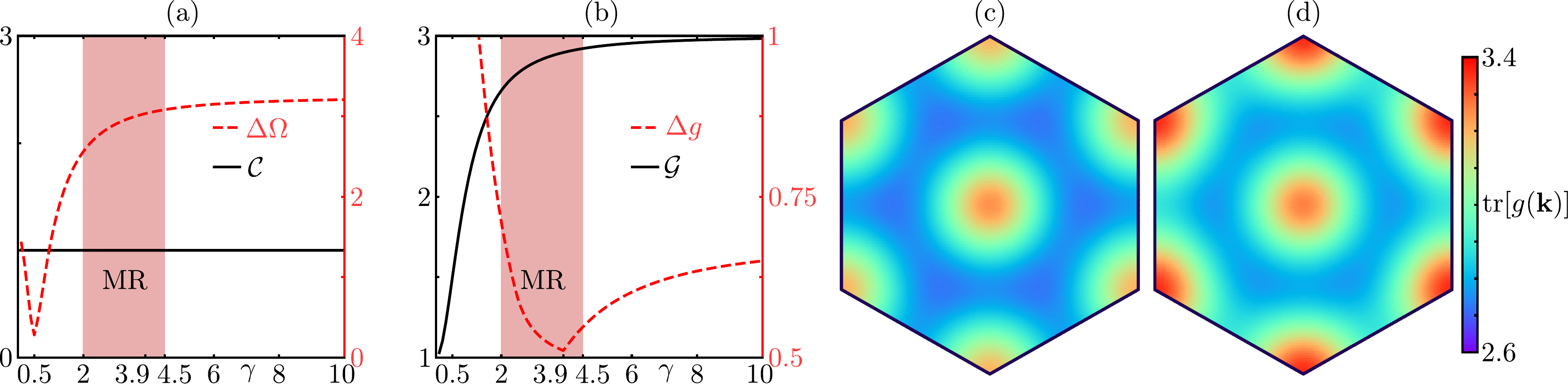}
\caption{(a) The total Berry curvature $\mathcal{C}=\frac{1}{2\pi}\int d\mathbf{k}^2 \Omega(\mathbf{k})$ (black solid line) and the fluctuation of Berry curvature $\Delta\Omega=\max[\Omega(\mathbf{k})]-\min[\Omega(\mathbf{k})]$ (red dashed line) as a function of the interlayer coupling $\gamma$. (b) The total quantum metric $\mathcal{G}=\frac{1}{2\pi}\int d\mathbf{k}^2{\rm tr}g(\mathbf{k})$ (black solid line) and the fluctuation of quantum metric $\Delta g=\max[{\rm tr}g(\mathbf{k})]-\min[{\rm tr}g(\mathbf{k})]$ (red dashed line) as a function of the interlayer coupling $\gamma$. Here, the shaded region indicates the presence of MR states. (c) and (d) are the quantum metric distribution over the moir\'e Brillouin zone for $\gamma=3.5\ {\rm eV}$ and $5.5\ {\rm eV}$, respectively. Here, we use a momentum grid of $100^2$, and $\Delta\mathcal{G}$ is renormalized over the momentum grid.
\label{fig:quantum_metric}
}
\end{figure}

In the context of fractional Chern insulators, quantum metric is often used as a heuristic measure of the stability of such phases. 
Here, we provide a detailed analysis of the quantum metric in the targeted flat band of the double TBG model. As shown in Fig.~\ref{fig:quantum_metric}(a) and (b), when $\gamma$ is zero, the system reduces to chiral twisted bilayer graphene (TBG), which can host Laughlin-like states. On the other hand, as $\gamma$ approaches infinity, the total quantum metric becomes 3 while the band Chern number remains $\mathcal{C}=1$, resembling the first Landau level, expected by recent papers on non-Abelian Moore-Read states~\cite{mr_Aidan, mr_Wang, mr_Ahn, mr_Xu, mr_zhang}.

Interestingly, the fluctuations of Berry curvature and quantum metric exhibit markedly different behaviors, while Berry curvature fluctuation reaches its minimum at small $\gamma$, quantum metric fluctuation is minimized at intermediate coupling strengths. Numerical simulations reveal the emergence of MR states only in the regime of intermediate coupling. In this context, compared to Berry curvature fluctuation, the weak quantum metric fluctuation serves a more reliable single-particle indicator for the emergence of MR states. 

Moreover, Fig.~\ref{fig:quantum_metric}(b) indicates that, although the total quantum metric may reach $3$ at extremely large coupling, this does not necessarily imply that the flat band mimics the first LL. Instead, the non-uniform quantum metric could alter the underlying physics~\cite{PhysRevLett.124.106803}. In this sense, we need to make a compromise between approaching $\mathcal{G}=3$ and reducing quantum metric fluctuation. 

 We next provide the quantum metric distributions for $\gamma=3.5\ {\rm eV}$ and $\gamma=5.5 \ {\rm eV}$ (see  Fig.~\ref{fig:quantum_metric}(b) and (c)), respectively. At intermediate coupling, the quantum metric is rather uniform. However, quantum metric at large coupling peaks at $\Gamma$ and $K$ points.

\section{single-hole energy}

\begin{figure}
\centering
\includegraphics[width=0.5\linewidth]{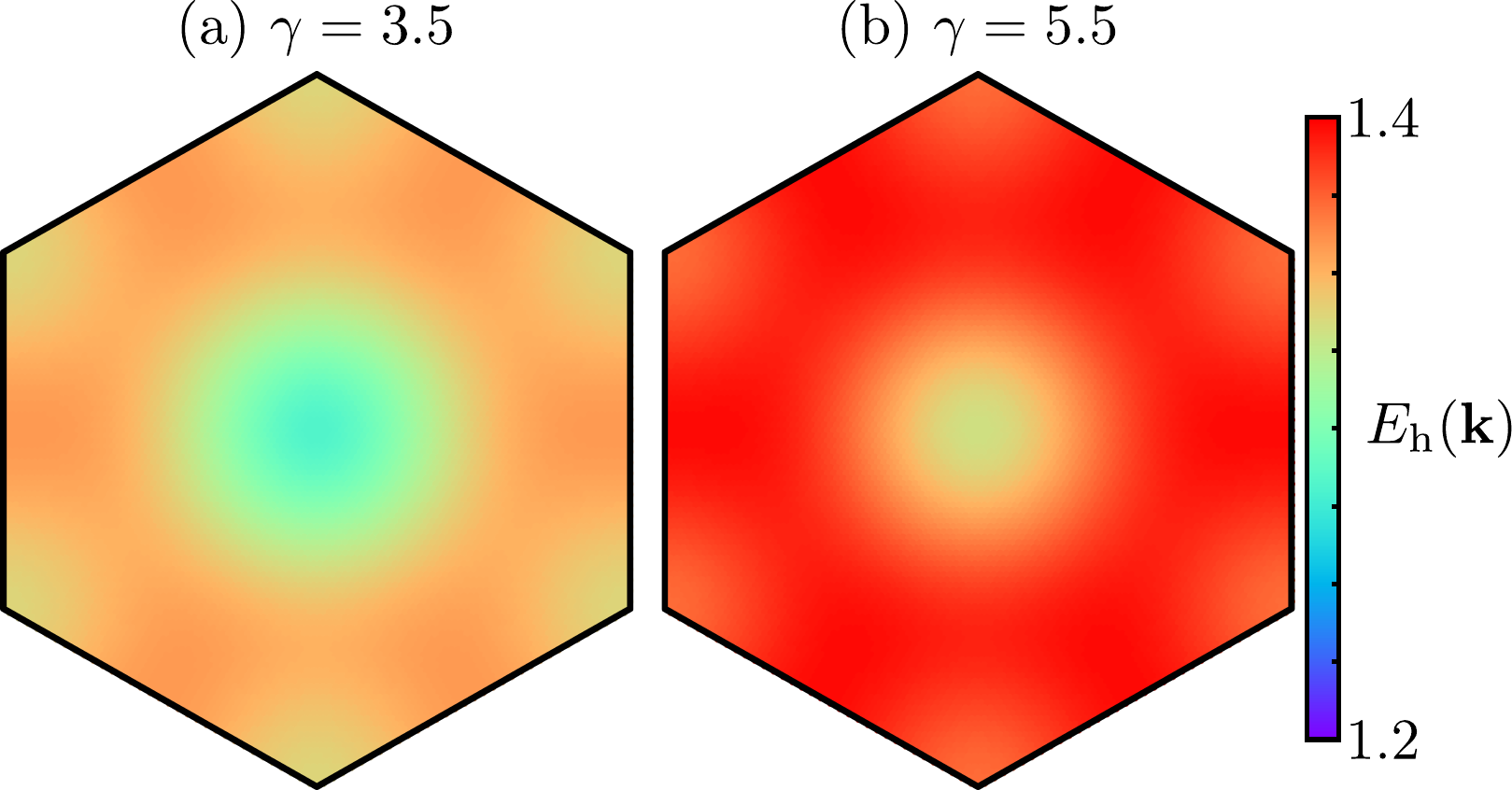}
\caption{Single-hole energy $E_{\text{h}}(\mathbf{k})$ in the moiré Brillouin zone for $\gamma=3.5$eV (a) and $\gamma=5.5$eV, respectively. 
\label{fig:single_hole_energy}
}
\end{figure}

In this section, we compare the single-hole energy in Landau levels and moiré minibands. For Landau levels, the single-hole energy can be shown analytically to be independent of the center of mass momentum. In contrast, for moiré minibands, the single-hole energy, in general, exhibits momentum dependence. 

We compute single-hole energy $E_{\text{h}}(\mathbf{k})= \sum_{\mathbf{k}'} \big( V_{\mathbf{k}'\mathbf{k}\mathbf{k}'\mathbf{k}} +  V_{\mathbf{k}\mathbf{k}'\mathbf{k}\mathbf{k}'} - V_{\mathbf{k}\mathbf{k}'\mathbf{k}'\mathbf{k}}-V_{\mathbf{k}'\mathbf{k}\mathbf{k}\mathbf{k}'}\big)$. The standard second quantization tells us 
\begin{eqnarray}
V_{\mathbf{k}_1 \mathbf{k}_2 \mathbf{k}_3 \mathbf{k}_4}=\frac{1}{2}\sum_{\bf q} V(\mathbf{q})F_{\mathbf{k}_1 \mathbf{k}_4}(\mathbf{q})F_{\mathbf{k}_2 \mathbf{k}_3}(-\mathbf{q}). 
\end{eqnarray}
Here  
\begin{eqnarray}
    F_{\mathbf{k}_1 \mathbf{k}_4}(\mathbf{q}) = \int d^2\mathbf{r}\varphi^{*}_{\mathbf{k}_1}(\mathbf{r})\varphi_{\mathbf{k}_4}(\mathbf{r})e^{i\mathbf{q}\cdot \mathbf{r}},~
        F_{\mathbf{k}_2 \mathbf{k}_3}(-\mathbf{q}) = \int d^2\mathbf{r}\varphi^{*}_{\mathbf{k}_2}(\mathbf{r})\varphi_{\mathbf{k}_3}(\mathbf{r})e^{-i\mathbf{q}\cdot \mathbf{r}},
\end{eqnarray}
and $\varphi_{\mathbf{k}}(\mathbf{r})$ is the single-particle wave function of either a Landau level or a Chern band. We also assume that the interaction potential is isotropic such that its Fourier transform $V({\bf q})$
only depends on $|{\bf q}|$. Because $F_{\mathbf{k}\mathbf{k}'}(\mathbf{q})=F_{\mathbf{k}'\mathbf{k}}^{*}(-\mathbf{q})$ and $\mathbf{q}\in\mathbb{R}^2$, we have
\begin{eqnarray}
V_{\mathbf{k}'\mathbf{k}\mathbf{k}'\mathbf{k}}&=&V_{\mathbf{k}\mathbf{k}'\mathbf{k}\mathbf{k}'}=\frac{1}{2}\sum_{\bf q}V(\mathbf{q})|F_{\mathbf{k}' \mathbf{k}}(\mathbf{q})|^2,\\
V_{\mathbf{k}\mathbf{k}'\mathbf{k}'\mathbf{k}}&=&V_{\mathbf{k}'\mathbf{k}\mathbf{k}\mathbf{k}'}=\frac{1}{2}\sum_{\bf q}V(\mathbf{q})F_{\mathbf{k} \mathbf{k}}(\mathbf{q})F_{\mathbf{k}' \mathbf{k}'}(-\mathbf{q}).
\end{eqnarray}

The $n$th Landau level (LL) on the $L_x\times L_y$ torus has an analytical single-particle wave function 
\begin{eqnarray}
    \varphi_{k}^{(n)}(x,y)=\frac{1}{(\sqrt{\pi} L_y \ell)^{1/2}}\sum_{m=-\infty}^{+\infty} e^{i\frac{2\pi y}{L_y}(k+mN_\phi)}e^{-\frac{1}{2\ell^2}\left[x-\frac{2\pi \ell^2}{L_y}(k+mN_\phi)\right]^2}H_n\left(\frac{x}{\ell}-\frac{2\pi\ell}{L_y}(k+mN_\phi)\right),
\end{eqnarray}
where $k$ is now the LL orbital index under the Landau gauge, $N_\phi$ is the number of magnetic flux, $\ell$ is the magnetic length, and $H_n$ is the Hermite polynomial. Then one can obtain~\cite{Yoshioka1983}
\begin{eqnarray}
F_{k_1k_2}^{(n)}(\mathbf{q}) =\delta_{q_x,\frac{2\pi s}{L_x}}\delta_{q_y,\frac{2\pi t}{L_y}}\tilde{\delta}_{k_1-k_2,t}L_{n}\left(\frac{|\mathbf{q}|^2 \ell^2}{2}\right)e^{-\frac{1}{4}|{\bf q}|^2\ell^2}e^{-i\frac{\pi s}{N_\phi}(2k_1+t)},
\end{eqnarray}
where $s$ and $t$ are integers, and $\tilde{\delta}$ is the periodic Kronecker delta function with period $N_\phi$.
Finally, we can get
\begin{eqnarray}
\label{Vkkkk}
         V^{(n)}_{k' k k' k} &=& \frac{1}{2}\sum_{\mathbf{q}} V({\bf q})\delta_{q_x,\frac{2\pi s}{L_x}}\delta_{q_y,\frac{2\pi t}{L_y}}\tilde{\delta}_{k'-k,t} e^{-\frac{1}{2}|\mathbf{q}|^2\ell^2}\left[L_{n}\left(\frac{|\mathbf{q}|^2\ell^2}{2}\right)\right]^2\nonumber,\\
         V^{(n)}_{k k' k' k}& =& \frac{1}{2}\sum_{\mathbf{q}} V({\bf q})\delta_{q_x,\frac{2\pi s}{L_x}}\delta_{q_y,\frac{2\pi t}{L_y}}\tilde{\delta}_{t,0} e^{-\frac{1}{2}|\mathbf{q}|^2\ell^2}e^{-2\pi i s\frac{k-k'}{N_\phi}}\left[L_{n}\left(\frac{|\mathbf{q}|^2\ell^2}{2}\right)\right]^2.
\end{eqnarray}
From these two equations, it is clear that both $V^{(n)}_{k' k k' k}$ and $V^{(n)}_{k k' k' k}$ only depend on $k-k'$ for the $n$th LL. So the hole dispersion, which requires summing up Eq.~(\ref{Vkkkk}) over $k'$, is independent of $k$. This constant $E_{\text{h}}(k)$ preserves the particle-hole symmetry in any Landau level. 

In moiré systems, since the Bloch states do not have the same analytic structure as Landau level orbitals in general, $V_{\mathbf{k'kk'k}}$ and $V_{\mathbf{k'kkk'}}$ then retain $\mathbf{k}$-dependent even after summing over $\mathbf{k}'$. To provide an intuitive picture, in Fig.~\ref{fig:single_hole_energy}, we numerically evaluate the single hole energy in the moiré Brillouin zone, in both the intermediate and strong coupling $\gamma$, $E_{\text{h}}$ clearly has a center-of-mass momentum dependence, which thus break the particle-hole symmetry. 
We note that since the fluctuation of the $E_{\text{h}}(\mathbf{k})$ is relatively small with intermediate coupling $\gamma$, the particle-hole symmetry is weakly broken, which is consistent with the discussion in the main text.

\begin{figure}
\centering
\includegraphics[width=0.8\linewidth]{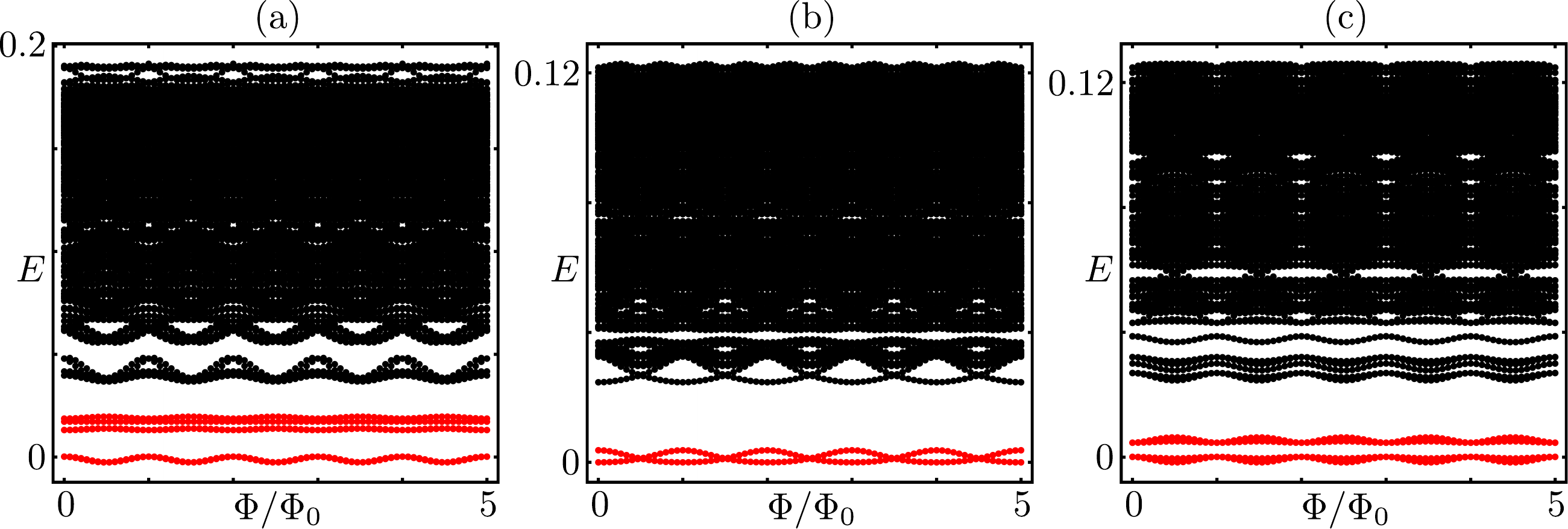}
\caption{Panels (a-c) are the spectral flow for MR states in the clusters with $20$, $26$, and $28$ sites, respectively. Here $\Phi_0$ is the flux quantum. For all plots, we employ a bare Coulomb interaction and $\gamma=3.5\ {\rm eV}$.
\label{fig:mr_flux}
}
\end{figure}

\section{Spectral flow for Moore-Read states}
In the main text, we have shown the gapped energy spectrum for MR states. Here we present the spectral flow for different system sizes by inserting an external magnetic flux. As shown in Fig.~\ref{fig:mr_flux}, the MR ground states evolve smoothly into one another while maintaining a visible gap from the excited states.

\section{Chiral graviton spectral function}
To further distinguish the Pfaffian and anti-Pfaffian states, we here employ the chiral graviton spectral function~\cite{form_factor_ll, graviton0, graviton00}.  
This method has been recently introduced to moiré FCI, where the graviton operator is now defined as~\cite{graviton1, graviton2, graviton3},  
\begin{eqnarray}
    O_{\pm}=\sum_{\mathbf{q}}\mathcal{D}_{\pm}(\mathbf{q})V(\mathbf{q})\rho_{\mathbf{q}}\rho_{\mathbf{-q}}.
\end{eqnarray}
Here, $\rho_{\mathbf{q}}$ is the projected density operator and $V(\mathbf{q})$ is the bare Coulomb interaction. The factor $\mathcal{D}_{\pm}(\mathbf{q})$ encodes a structure consistent with the $C_{3v}$ lattice symmetry and translational invariance of the 2D triangular lattice (as relevant to small angle twisted materials), and reduces to $(q_x\pm iq_y)^2$ in the long-wavelength limit. 
The chiral graviton spectral function can then be defined as 
\begin{eqnarray}
    I_{\pm}=\sum_n|\langle \psi_n|O_{\pm}|\psi_0\rangle|^2\delta (E-E_n+E_0),
\end{eqnarray}
with $E_0$ ($E_n$) and $|\psi_0\rangle$ ($|\psi_n\rangle$) being the ground state ($n$-th excitation state) energy and eigenvector.

\begin{figure}
\centering
\includegraphics[width=0.6\linewidth]{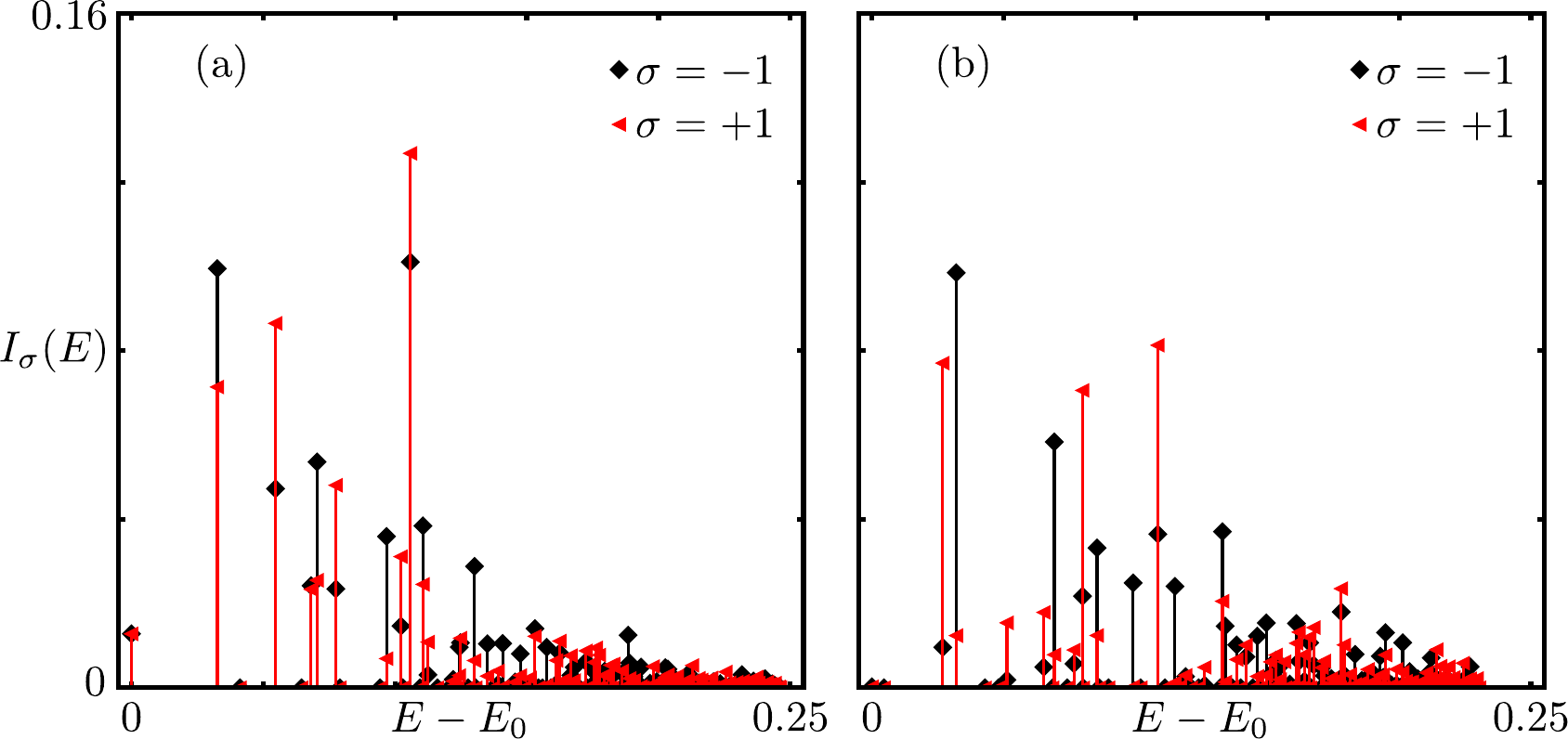}
\caption{Chiral graviton spectral function with $\mathcal{D}_{\pm}(\mathbf{q})$ respecting the crystal symmetry of moiré lattice as a function of energy for $13$ electrons (a) and $14$ electrons (b), respectively. Here, all data points are normalized by $I_0=\frac{1}{2}\langle\psi_0|(O^\dagger_{+}O_{+}+O^\dagger_{-}O_{-})|\psi_0\rangle$
\label{fig:graviton_crystal_potential}
}
\end{figure}

With this, we implement the graviton spectra with both $26$site-cluster and $28$site-cluster. Our results are shown in Fig.~\ref{fig:graviton_crystal_potential}. For both even and odd number of electrons, we found that the graviton spectra for positive and negative chiralities has nearly equal weight, which suggests that the ground state is a mixed state of Pfaffian and anti-Pfaffian.

\section{Composite Fermi liquid at weak coupling}
\label{CFL}

\begin{figure}
\centering
\includegraphics[width=0.8\linewidth]{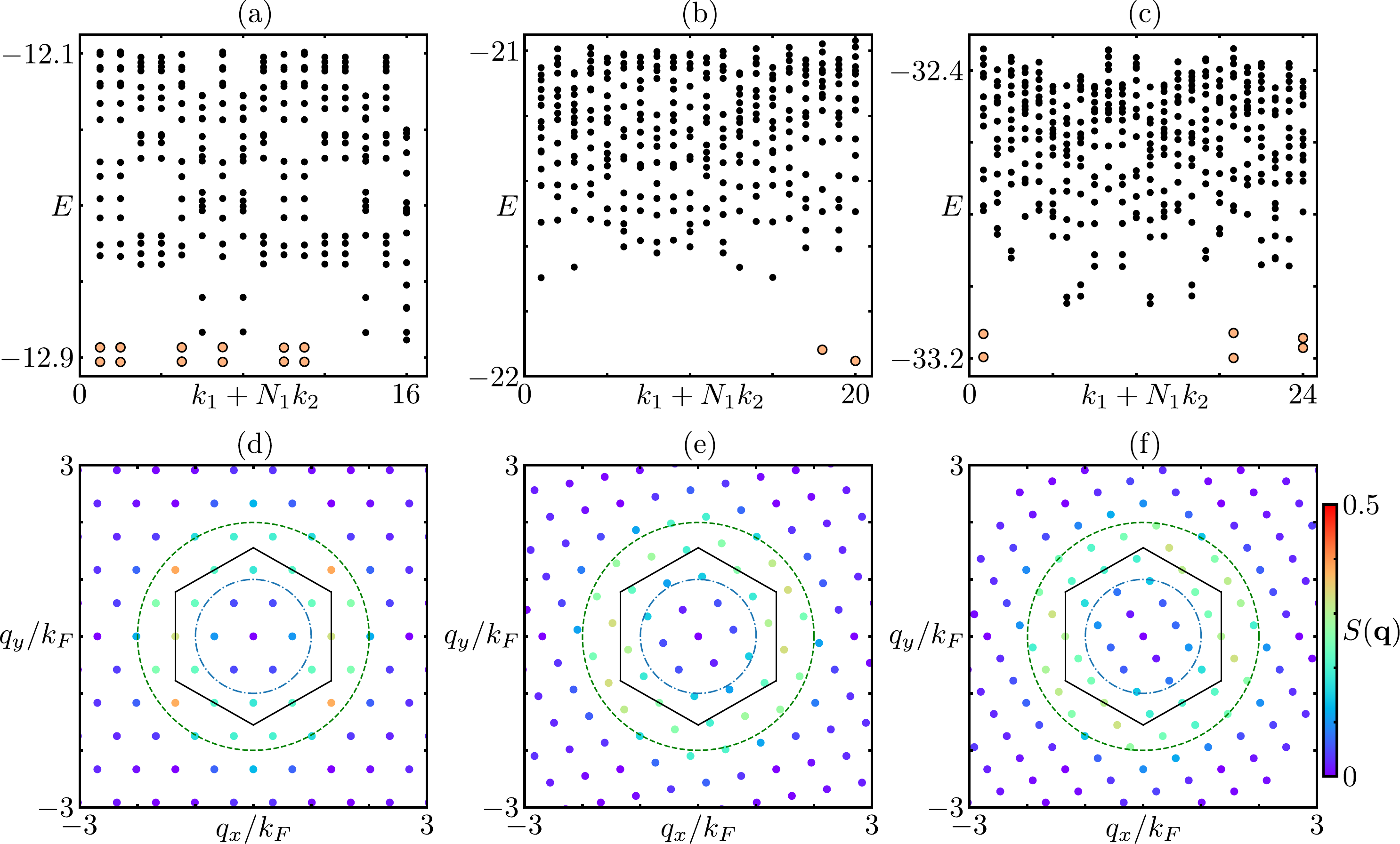}
\caption{(a-c) are the energy spectra for system size $4\times 4$, $4\times 5$, and $4\times 6$, respectively. The yellow dots indicate the ground states for composite Fermi liquid. (d-f) are their corresponding structure factors, respectively. Here, the solid, dashed, and dashdot lines represent the boundary of moir\'e Brillouin zone, $|\mathbf{q}|=k_F$, and $|\mathbf{q}=|2k_F$, respectively.
For all plots, we use $\gamma=0$ and a bare Coulomb interaction.
\label{fig:cfl}
}
\end{figure}

\begin{figure}
\centering
\includegraphics[width=0.3\linewidth]{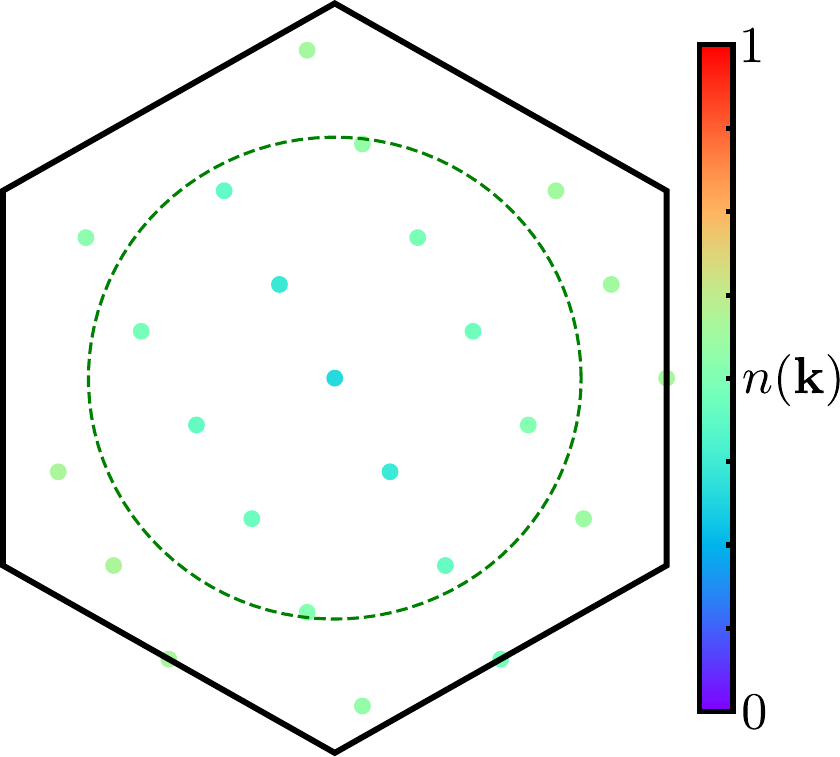}
\caption{The electron occupation $n(\mathbf{k})$ of the CFL ground state in the moiré Brillouin zone. Here, the dashed line indicates $|\mathbf{k}|=k_F$. 
\label{fig:cfl_nk}
}
\end{figure}

In the main text, we have mentioned that the emergence of the composite Fermi liquid at weak coupling strength originates from the lowest LL physics in the chiral limit of the TBG. We now display the result in Fig.~\ref{fig:cfl}. 
Our first evidence is the low-lying energy spectrum, where the ground states in various system sizes show one-to-one correspondence with the recent observations in tMoTe$_2$ and the lowest LL at half-filling~\cite{PhysRevLett.131.136502, PhysRevLett.131.136501}. 
Another feature of the composite Fermi liquid is the singularity at $|\mathbf{q}|=2k_F$, associated with the backscattering process around the Fermi surface at $k_F$ with Coulomb interactions~\cite{composite_fermion_science}. 
To observe this, we employ a structure factor calculation. As shown in Fig.~\ref{fig:cfl}(d-f), we found that the structure factor becomes pronounced around $2k_F$ and delays rapidly when further increasing $|\mathbf{q}|$, consistent with a singularity at $2k_F$. Here $A_{BZ}$ is the area of the moiré Brillouin zone.
And by calculating the electron occupation $n(\mathbf{k})$, we found that in the electron picture, there is no obvious jump near the Fermi surface at $k_F$ (see Fig.~\ref{fig:cfl_nk}).

\section{PES of charge density waves}

For the particle-cut entanglement spectrum (PES), we divide the system into $N_A$ and $N_B=N_e-N_A$ particles, with $N_e$ the total number of electrons. The PES is defined as the eigenvalues $\{\xi\}$ of $-\ln\rho_{A}$, where $\rho_A=\text{tr}_{B}[\frac{1}{N_d}\sum_{i=1}^{N_d}|\Psi_i\rangle\langle \Psi_i|]$ is the reduced density matrix of the subsystem $A$, with $|\Psi_i\rangle$ the corresponding ground states and $N_d$ the ground-state degeneracy. As we effectively create holes in the system by tracing out $B$, the PES contains the information of quasihole excitations, serving as a fingerprint of either FCIs or CDWs.

In the main text, we have presented the PES of the CDW phase for both the $20$site-cluster and $26$site-cluster with $N_A=4$. 
Here, we extend our analysis to larger system sizes and various values of $N_A$ to further demonstrate the stability of the CDW phase at large $\gamma$. 

\begin{figure}
\centering
\includegraphics[width=0.4\linewidth]{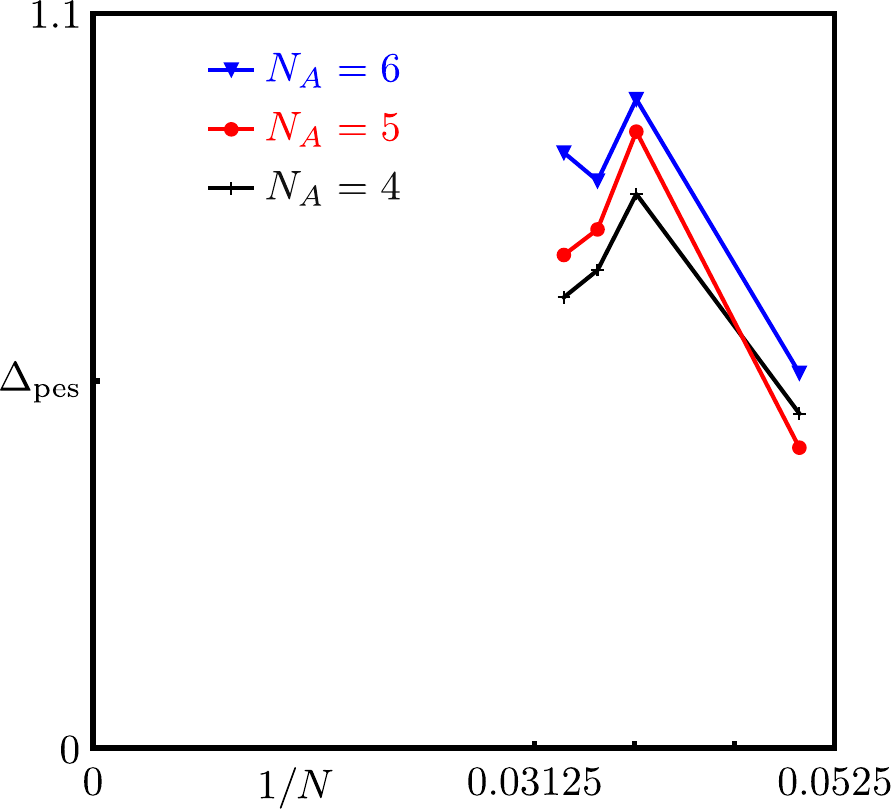}
\caption{Particle-cut entanglement spectrum gap for CDW ground states as a function of system sizes $N$ with different $N_A$ (the number of particles in A subspace). Here, we employ $\gamma=5.5$eV with $20$, $26$, $28$, and $30$site-clusters, respectively. 
\label{fig:pes_scaling}
}
\end{figure}

As shown in Fig.~\ref{fig:pes_scaling}, we further include $28$site-cluster and $30$site-cluster, and increase $N_A$ up to $6$. The PES gap for CDWs remains open across all these cases and shows no tendency toward closing, providing strong evidence for the robustness of the CDW phase in the large-$\gamma$ regime.

\section{charge density wave at larger system sizes}
\begin{figure}
\centering
\includegraphics[width=0.6\linewidth]{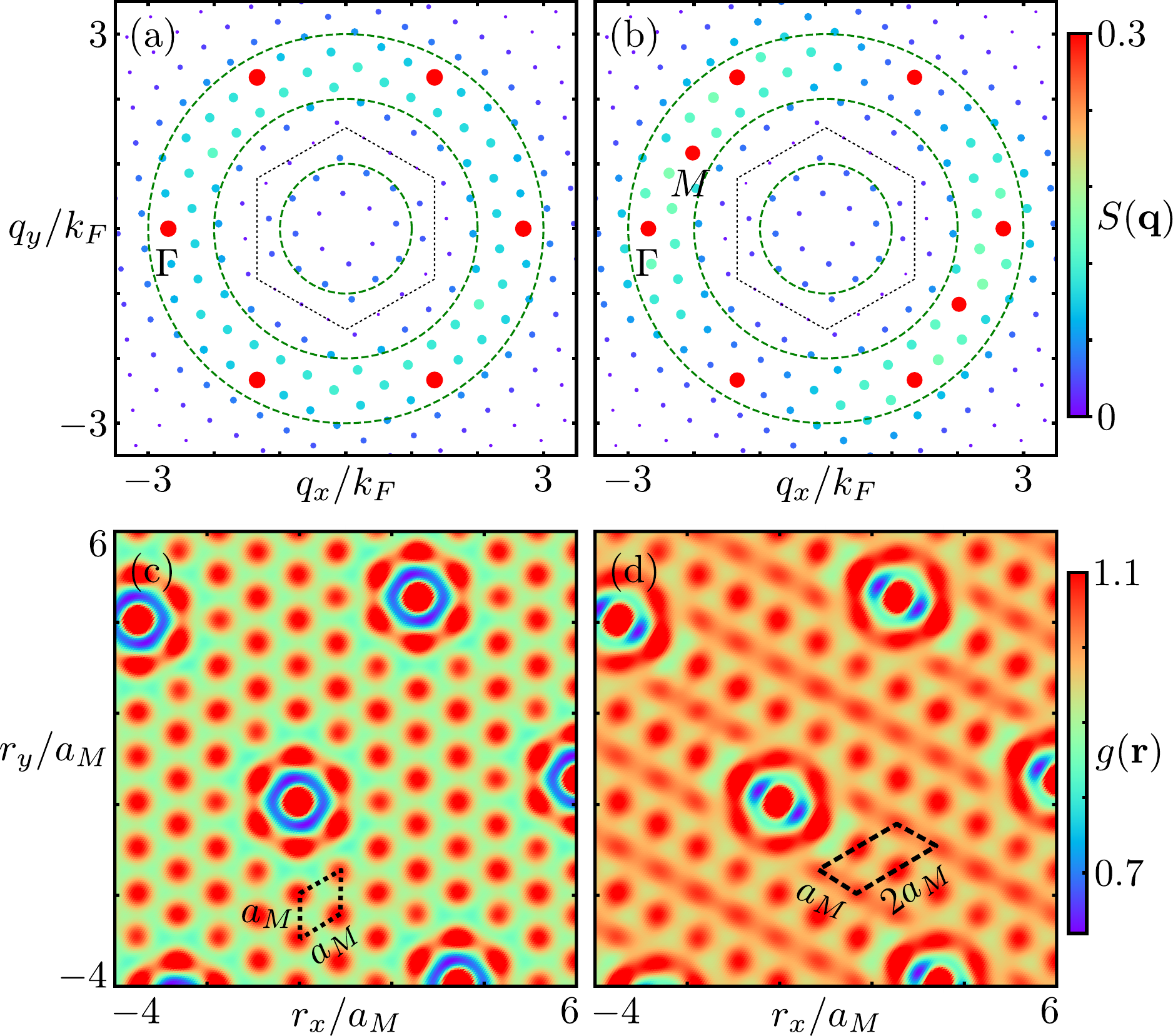}
\caption{(a) and (b) are the structure factors of ground states in the MR ($\gamma=3.5\ {\rm eV}$) and CDW order ($\gamma=5.5\ {\rm eV}$) region for a $30$-site system, respectively. (c) and (d) are their corresponding pair correlation functions. In (c), the big red dots originate from the divergence of structure factor at $\Gamma$ points, which only reflects the moir\'e potential. In comparison, (d) reveals an extra $1\times 2$ crystalline structure, which arises due to the structure factor peaks at $M$ points. Here the dotted and dashed rhombus in pair correlation function plots indicate the $1\times 1$ and $1\times 2$ unit cell configuration, respectively. 
\label{fig:cdw_30}
}
\end{figure}

In this section, we present evidence for charge density wave (CDW) orders in a system comprising 30 sites. As shown in Fig~\ref{fig:cdw_30}, when the system is in the Moore-Read (MR) state, peaks appearing at $\Gamma$ points in the structure factor which only reflect the moiré potential, thus beside $1\times 1$ unit cell configurations for moiré potential, there is no crystalline order in the pair correlation function. By contrast, when a CDW order is present, extra peaks appear at $M$ points. Compared to the case of MR states, an extra crystalline order emerges with a $1\times 2$ unit cell configurations in pair correlation function plot.

We note that the structure peaks at the $\Gamma$
points are omitted in Fig.~2 of the main text, as they only reflect the moiré potential and do not provide additional insights. 
Here, we include them to facilitate a comparison with the contribution from the structure factor peaks at $M$ points. 

\section{energy gap scaling}

\begin{figure}
\centering
\includegraphics[width=0.8\linewidth]{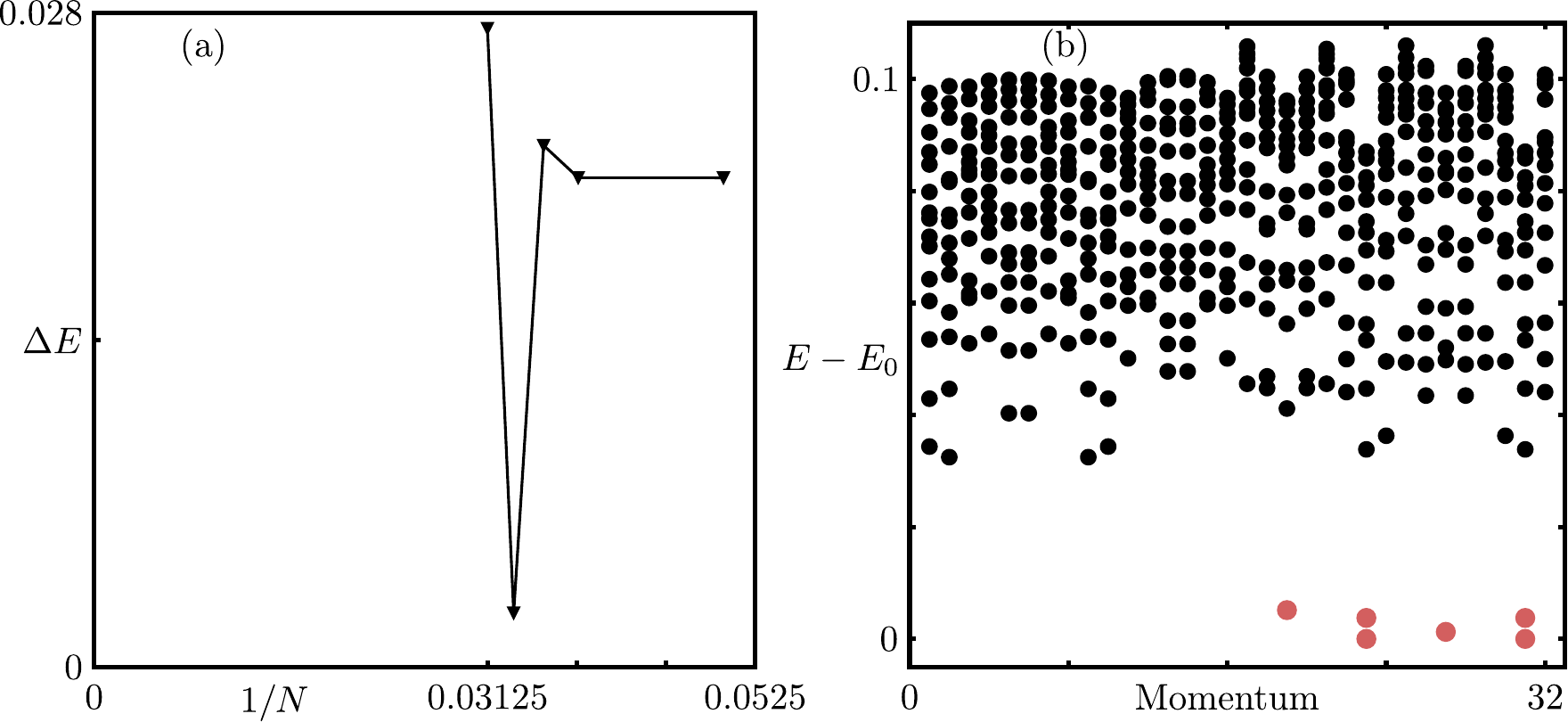}
\caption{(a) The energy gap between MR states and higher excitation states as a function of system sizes $N$. Here, we employ $\gamma=3.5$eV with $N=20$, $26$, $28$, $30$, and $32$site-clusters, respectively. (b) The corresponding energy spectrum for $32$site-clusters. Here the red dots indicate MR ground states with correct center-of-mass momentum.  
\label{fig:gap_scaling}
}
\end{figure}

In Fig.~\ref{fig:gap_scaling}(a), we plot the excitation gap versus inverse system size $1/N$ (or $1/2N_e$) for MR states. From $20$site to $28$site clusters, the gap remains essentially constant, indicating good size convergence. 
In the $30$-site cluster, however, the gap drops quickly. We attribute this reduction to the specific cluster geometry: despite the smaller gap, the particle-cut entanglement spectrum shown in the main text still exhibits a sizable entanglement gap, suggesting that the underlying phase remains robust. We therefore expect that the gap size will recover in clusters with a different tiling and remain open in the thermodynamic limit. 
Align with this, we further perform calculations on $32$site-cluster, where the low-lying energy spectrum again shows $6$ quasi-degenerate states at the correct center-of-mass momentum for MR states, with a larger gap separating from the excitation states (see Fig.~\ref{fig:gap_scaling}(a) and (b)).

\section{quasihole counting for Moore-Read states and charge density waves}

Here, we detail how to derive the quasihole counting of the Moore-Read states. In the thin-torus limit, the ground states of non-Abelian fractional topological states at band filling $\nu=k/(kM+2)$ should fulfill the rules that, any $kM+2$ consecutive sites host exactly $k$ particles, and the distance between two particles is at least $M$~\cite{Ardonne_2008}. 
For fermionic Moore-Read states at half-filling, 
 where $k=2$ and $M=1$, this yields ground states exhibiting a sixfold degeneracy for an even number of electrons and a twofold degeneracy for an odd number of electrons. 
A quasihole state is then characterized by one string of $(kM+2)$ consecutive sites carrying particles less than $k$. 
For the quasihole counting, it is just the number of how many such states are allowed when putting $N_e$ particles in $N$ sites with $N_e/N<\nu$. 

We give a concrete example below. For the Moore-Read quasihole states with four electrons in $N$ sites ($N>8$), we categorize all possible configurations into three groups:
\begin{itemize}
    \item No $0110$ strings. The allowed configurations are the combinations of 4 strings of $10$ and zeros,
    \begin{eqnarray}
        n_1=N\cdot\frac{(N-8+4-1)!}{(N-8)!\cdot 4!}.
    \end{eqnarray}
    \item One $0110$ string. The permissible configurations are the combinations of one string of $01100$, two strings of $10$, and zeros,
        \begin{eqnarray}
        n_2=N\cdot\frac{(N-9+3-1)!}{(N-9)!\cdot 2!}.
    \end{eqnarray}
    \item Two $0110$ strings. The allowed configurations are the combinations of two strings of $0110$ and zeros,
        \begin{eqnarray}
        n_3=N\cdot\frac{(N-8+2-1)!}{(N-8)!\cdot 2!}.
    \end{eqnarray}
\end{itemize} 
By summing up all these possibilities, we get the quasihole counting for four electrons, which is expected to equal the PES counting with $N_A=4$. For $N=20$, $24$, $28$, and $30$, $n=n_1+n_2+n_3=3965$, $9282$, $18571$, and $25185$, respectively, all of which match the numerics. 
We note that, including more particles would make the quasihole counting more complicated, thus precluding to find an explicit formula for the PES counting with a general $N_A$ and $\nu$~\cite{PhysRevB.73.245334}.

For the charge density wave case, the ground state is characterized by a fixed configuration, then the number of quasihole states is just the possibilities of putting $N_A$ particles into $N_e$ sites ($N_e$ is the total number of electrons). We then have 
\begin{eqnarray}
    n=d\begin{pmatrix}
        N_A\\N_e
    \end{pmatrix}.
\end{eqnarray}
Here, $d$ is the number of charge ordered states. In the main text, for $N_A=4$ and $N_e=10$, we have $n=420$, which indicates that $d=2$ states are characterized by charge density waves. 

\end{document}